\documentclass[12pt]{article}
\usepackage{graphicx}
\usepackage{epsfig}
\usepackage{amsmath}
\usepackage{amsthm}
\usepackage{amssymb}

\usepackage[colorlinks,citecolor=blue,urlcolor=blue,backref=page,backref=page]{hyperref}

\usepackage[round]{natbib}
\usepackage{color}

\usepackage{units}

\newcommand{\n}[1] {\mbox{\boldmath{$#1$}}}

\newtheorem*{example}{Example}

\newtheorem{res}{Result}
\newtheorem{pro}{Proposition}
\newtheorem{defn}{Definition}
\newcommand{\be}{\begin{eqnarray}}
	\newcommand{\ee}{\end{eqnarray}}
\newcommand{\beq}[1]{\begin{equation}\label{#1}}
	\newcommand{\eeq}{\end{equation}}
\newcommand{\ba}{\begin{eqnarray*}}
	\newcommand{\ea}{\end{eqnarray*}}

\newcommand{\diid}{\mathop{\sim}\limits^{\mathrm{iid}}}
\newcommand{\erre}{\mbox{I$\!$R}}

\title{Model Uncertainty and Missing Data: An Objective Bayesian Perspective}
\author{Gonzalo Garc\'ia-Donato$^1$ \and María Eugenia Castellanos$^2$ \and Stefano Cabras$^3$ \and Alicia Quirós$^4$ \and Anabel Forte$^5$}
\date{
	\footnotesize{$^1$ Universidad de Castilla La Mancha (Spain); $^2$ Universidad Rey Juan Carlos (Spain); $^3$ Universidad Carlos III (Spain), $^4$ Universidad de León (Spain); $^5$ Universidad de Valencia (Spain)}\\
September, 2024}
\begin{document}
	\maketitle

\begin{abstract}
The interplay between missing data and model uncertainty—two classic statistical problems—leads to primary questions that we formally address from an objective Bayesian perspective. For the general regression problem, we discuss the probabilistic justification of Rubin's rules applied to the usual components of Bayesian variable selection, arguing that prior predictive marginals should be central to the pursued methodology. In the regression settings, we explore the conditions of prior distributions that make the missing data mechanism ignorable. Moreover, when comparing multiple linear models, we provide a complete methodology for dealing with special cases, such as variable selection or uncertainty regarding model errors. In numerous simulation experiments, we demonstrate that our method outperforms or equals others, in consistently producing results close to those obtained using the full dataset. In general, the difference increases with the percentage of missing data and the correlation between the variables used for imputation. Finally, we summarize possible directions for future research.
\end{abstract}

\noindent%
{\it Keywords:}  Bayes factor; $g$-priors; Ignorability; Objective prior distribution; Rubin's rules
\vfill

\newpage

\section{Introduction}
Model uncertainty is a broad term for situations where a true data-generative model is assumed unknown. Paradigmatic model uncertainty problems include model choice, hypothesis testing, variable selection (VS), and model averaging. From a Bayesian perspective, a formal tool for addressing such problems is the posterior distribution over the model space. It assigns, conditionally on the data, the probability of each model and constitutes a comprehensive tool that is the basis for addressing \emph{all} types of questions in model uncertainty scenarios.

Obtaining the posterior distribution, from straight probability arguments, entails severe difficulties of quite a different nature, particularly from an objective perspective \cite[see][for a detailed discussion of objectivism in Bayesian statistics]{Ber06}. Many of these challenges have to be with the conditions that prior distributions must satisfy for the Bayes factors (BF) to be well-defined \citep{Jeff:61, Kass:Raft:95, Berg:Peri:01}, multiplicity issues \citep{Scott:Berg:05}, and numerical problems (not only for the computation of the marginal distribution for each model but also for sampling strategies when the model space is very large). Motivated by these, the field of model uncertainty has received considerable attention in recent decades and has acquired high levels of maturity (see \cite{Baetal11} for a pioneering attempt to standardize good practices and \cite{TaVa22} for a collection of contemporaneous techniques in the field). Unfortunately, for problems with missing data, many of the proposed solutions do not apply directly, and their bases must be carefully reconsidered. A distinguished case is that of $g$-priors \citep{Zell:86}, and the enormous number of generalizations they inspired \cite[see, for example,][for an extremely popular reference]{liang08}, which dependence on a complete fixed design matrix makes them useless in the case of missing observations. 


\subsection{Goals and structure}    
In this study, we focus on deriving reliable objective posterior distributions for model uncertainty with missing data in light of the standards in \cite{Baetal11} and all the references there compiled. 
For this task, Section~\ref{sec.initial} presents the problem from a broad perspective, emphasizing the interplay between Rubin's rules and posterior model probabilities. 

The remainder of the article is organized as follows. Section~\ref{sec.reg} derives several equivalent expressions for the prior predictive marginals (the key ingredients of the posterior model probabilities and the BF) in regression models. We then establish the conditions for the ignorability of the missing mechanism and propose numerical strategies for marginal computation using simulation methods. We conclude this section with general considerations regarding the assignment of objective prior distributions on model parameters. Section~\ref{sec.lm} derives a complete methodology for VS in linear models with Gaussian errors based on a new prior distribution that extends $g$-priors, in a way that the dependence on the missed values of covariates is circumvented. Section~\ref{sec.errors} addresses a non-nested situation that has received very little attention in the literature. The considered linear models disagree on the distribution of errors, which we illustrate by comparing the different forms of the error covariance matrix. Section~\ref{sec.example} evaluates the performance of the model uncertainty procedure and compares it with results obtained with the fully observed dataset (oracle method); listwise deletions and some procedures proposed in the literature. To this purpose we employ several simulated and real datasets with varying levels of missing data. Finally, Section~\ref{sec:conclusions} concludes the paper by describing several directions for future research. 
    
\subsection{A brief review of the literature}\label{subsec.literature}
Research on imputation methods \emph{per se} is a classical topic in Bayesian literature on missing data. Recent studies on this topic include those of \cite{Danetal16, Mosetal20, Go20} and \cite{Abetal22}. Other projects have focused on estimating a fixed model in the presence of missing data without considering model uncertainty, such as \cite{Ibraetal02} for generalized linear models, \cite{Eretal16} for longitudinal models, and \cite{Er19} for epidemiologic modelization with time-varying covariates.

Within model uncertainty problems, several Bayesian researchers have responded to the difficulties of obtaining a sensible posterior distribution in the presence of missing data by proposing alternative criteria for model selection. This is the path taken by \cite{Ibraetal02}, who introduced a new criterion similar to the BIC, and \cite{Ceetal06} and \cite{Ibetal06}, who extended DIC for missing data models or when missing data were present. \cite{cohen2021normalized} proposed a normalized version of AIC and BIC that allows the selection of variables without providing the full model uncertainty quantification based on the model posterior distribution. Similarly, \cite{Danetal12} proposed a model choice measure based on a posterior predictive distribution. However, these measures do not provide any uncertainty regarding the model selection question and have complicated interpretability.

In the context of VS, methods based on the direct use of a posterior distribution have been published by \cite{Biometrics2005}, \cite{Boetal20}, and \cite{Storetal20}. The last two emphasize imputation methods with specific applications in mind. 
By contrast, \cite{Biometrics2005} is more general and has developed a full methodology to define and implement the computation of posterior distributions for VS with missing data. 
These studies consist of excellent deployments of Bayesian machinery to impute missing data. However, aspects that govern the essential properties of the resulting methods in relation to their model uncertainty are essentially unnoticed. For instance, vague priors are used---despite the many warnings advising against it---and there is no discussion regarding multiplicity issues, thus increasing the chances of reporting false positives, which could be inadvertently caused by a casual choice of initial probabilities assigned over the model space. In Section~\ref{sec.example}, we reproduce the simulation experiment in  \cite{Biometrics2005} and show that their results are significantly outperformed by the posterior distribution we derive. 

One work that connects, in spirit, to ours is  \cite{Hoijtink_2019}. These authors argued that research on BF with missing data has received no attention in the literature and proposed easily implementable strategies to combine software for multiple imputation and BF calculations.
Their study limits to Bayesian testing; hence, it is strictly included in our study.

\section{Notation and posterior probabilities}\label{sec.initial}

\subsection{Notation}\label{sec:notation}
Following the convention in \cite{LitRub20}, let $\n d_{(0)}$ identify the available values in a dataset and let $\n d_{(1)}$ denote the missed observations. 
In model uncertainty there are several models under consideration, that we denote $\n\gamma$. This discrete parameter takes values from the set of possible alternatives $\Gamma$ (also called the model space), and its posterior distribution based on the available data is denoted by $p(\n\gamma\mid\n d_{(0)})$. We adopt the $\Gamma$-closed perspective and assume that one of the models in the model space is the true model.

Regarding the remainder of the notation, the letter $f$ denotes the density function for unknown but potentially observable random variables and vectors. The distribution of the parameters within each model, either a priori or a posteriori, is denoted as $\pi$ and the marginal distributions are labeled $m$. The particular form of any of these functions under a given model $\n\gamma$ is identified by the corresponding sub-index, and, for instance, $f_\gamma$ is the form of $f$ proposed under model $\n\gamma$.

\subsection{Rubin's rules and model posterior probabilities} \label{sec.RR}
What has been termed ``the key Bayesian motivation for multiple imputations'' \citep[][p.476]{Rubin96} is a simple probabilistic identity that has greatly influenced the area of statistical methods to handle missingness. In the context of model uncertainty, this can be written as
\begin{equation}\label{eq.post1}
	p(\n\gamma\mid \n d_{(0)})=\int p(\n\gamma\mid \n d_{(0)},\n d_{(1)})\, m(\n d_{(1)}\mid \n d_{(0)})\, \mbox{d}\n d_{(1)}
\end{equation}
where $p(\n\gamma\mid \n d_{(0)},\n d_{(1)})$ is the posterior probability of $\n\gamma$ given the completed dataset and $m(\n d_{(1)}\mid \n d_{(0)})$ is the posterior predictive distribution for $\n d_{(1)}$. 

This identity suggests a possible strategy for approximating $p(\n\gamma\mid \n d_{(0)})$ by creating multiple imputations of the dataset and then reporting the mean of the model’s posterior probabilities over the completed datasets. This procedure aligns with Rubin's rules and can be easily implemented with specific software for imputation \citep[such as {\tt mice} by][]{mice} properly combined with software for posterior model probabilities \cite[such as {\tt BayesVarsel} by][]{Gar-DonFor18}. However, a close examination of the posterior predictive distribution reveals that it depends on the posterior probability of the model, that is
\begin{equation}\label{marg.imput}
m(\n d_{(1)}\mid \n d_{(0)})=\sum_{\gamma\in\Gamma} m_{\gamma}(\n d_{(1)}\mid \n d_{(0)})\, p(\n\gamma\mid \n d_{(0)}).
\end{equation}
The fact that the target probability $p(\n\gamma\mid \n d_{(0)})$ appears on both sides of Equation~\eqref{eq.post1} hampers the formulation of the mentioned strategies, \emph{a la} Rubin's rules, simply because the distribution for imputation \eqref{marg.imput} is unknown. Unavoidably, a single model must be used for the imputation step (consciously or unconsciously), leading to a methodology that is not endorsed by the probabilistic equation in \eqref{eq.post1}. This is the basis of the ``Impute Then Select'' method in \cite{Biometrics2005}. 

Alternatively, as is routinely performed in estimation problems, we can envisage a Gibbs sampling algorithm, where $(\n\gamma, \n d_{(1)})$ are jointly drawn from their full conditional distributions. Such a strategy is feasible because within the conditional $\n d_{(1)}\mid\n\gamma$, the model is known. The problem is that the resulting method is computationally demanding especially when the cardinality of $\Gamma$ is moderate or large. This is because \emph{each} Gibbs sampling step requires handling a discrete space of enormous cardinality. This is the basis of the method  ``Simultaneously Impute And Select'' (SIAS) proposed in \cite{Biometrics2005}.

The starting point of our research follows directly from the application of Bayes' theorem:
\begin{equation}\label{eq.post2}
	p(\n\gamma\mid \n d_{(0)})=\frac{m_\gamma(\n d_{(0)})p(\n\gamma)}{\sum_{\gamma'\in\Gamma} m_{\gamma'}(\n d_{(0)})p(\n\gamma')}.
\end{equation}
 Obviously, \eqref{eq.post1} and \eqref{eq.post2} are equivalent, but the latter focuses on marginals---which are key quantities in our research---and, in principle, does not have an interpretation in terms of Rubin's rules.

\section{Model choice in regression settings with missing data}\label{sec.reg}
In a regression problem, the data consists on samples of size $n$ of a dependent variable $\n y$ and $p$ explanatory variables $\n x_1,\ldots,\n x_p$. We denote $X$ the matrix with entries $x_{ij}$. Of these, the elements that will conform $\n d_{(0)}$ are encoded in a binary matrix $M=(M_{ij})$ with dimensions $n\times (p+1)$. A value of $M_{ij}=1$ for $j\le p$ indicates that $x_{ij}$ is unavailable (to the analyst), whereas a value of $M_{ij}=0$ indicates that it is available. In this notation, column $p+1$ of $M$ represents the missingness of the response variable $\n y$. We denote $\widetilde{\n x}_{(0)}=\{\widetilde{x}_{ij}:\widetilde{M}_{ij}=0, \; \mbox{ for } 1 \leq i \leq n, \; 1 \leq j \leq p\}$ and $\widetilde{\n y}_{(0)}=\{\widetilde{y}_{i}:\widetilde{M}_{i, p+1}=0, \; \mbox{ for } 1 \leq i \leq n\}$ and a similar notation for ${\n y}_{(1)}$ and ${\n x}_{(1)}$. Hereafter, we use $(\n y_{(0)}, \n x_{(0)}, M)\equiv \n d_{(0)}$ and $(\n y_{(1)}, \n x_{(1)})\equiv \n d_{(1)}$.

\subsection{First considerations}\label{MURM}
%
We assume that models $\n\gamma\in\Gamma$ can be expressed as:
\begin{eqnarray}
f_\gamma(\n y,\n x_1,\ldots,\n x_p,M \mid \n\alpha,\n\beta_\gamma, \n \nu, \n \psi)
&=& f_\gamma(\n y\mid \n x_1,\ldots,\n x_p,\n\alpha,\n\beta_\gamma)\label{regres.gamma.1}\\
&\times&\prod_{i=1}^n\, f((x_{i1},x_{i2},\ldots,x_{ip})\mid \n\nu) \label{regres.gamma.2}\\
&\times& f(M\mid \n y, \n x_1,\ldots,\n x_p,\n\psi).\label{regres.gamma.3}
\end{eqnarray}

Notice that competing models only differ on \eqref{regres.gamma.1}, that is, on how the covariates influence $y$.
This conditional distribution is presented in a general manner to cover a wide range of situations and, in particular, the two that will be treated in detail in Section~\ref{sec.lm} (variable selection) and Section~\ref{sec.errors} (concerning competing models agreeing on the relevant covariates but disagreeing on the density).

In the introduction of the models, we have opted for a nonstandard notation in which all covariates appear in the conditioning. This does not necessarily imply an effective dependence of $y$ on \emph{all} $p$ covariates on all models in $\Gamma$ (usually certain models in the model space will not depend on any covariate (e.g., a model with only the intercept), and even certain covariates in the database will not appear in any competing model). The reason for this additional complexity becomes clear when the missing data problem develops. However, we anticipate that variables that do not appear in any competing model for $y$ are valuable in the imputation process \eqref{regres.gamma.2}. Regarding the regression parameters, we denote those appearing in all competing models (if any) as $\n\alpha$ (e.g., the intercept) and those specific to $\n\gamma$ (such as the regression parameter corresponding to an included variable) as $\n\beta_\gamma$.

Much of the literature assumes that the values of the covariates are known (either because the data come from a designed experiment or as a simplification), in which case, the marginal to be inserted in \eqref{eq.post2} would be
\begin{equation}\label{mgamma.complete}
	m_\gamma(\widetilde{\n y}\mid \widetilde{\n x}_1,\ldots,\widetilde{\n x}_p)=\int f_\gamma(\widetilde{\n y}\mid \widetilde{\n x}_1,\ldots,\widetilde{\n x}_p,\n\alpha,\n\beta_\gamma) \pi_\gamma(\n\alpha,\n\beta_\gamma\mid \widetilde{\n x}_1,\ldots,\widetilde{\n x}_p)\mbox{d}[\n\alpha,\n\beta_\gamma],
\end{equation}
where $\pi_\gamma(\n\alpha,\n\beta_\gamma\mid \widetilde{\n x}_1,\ldots,\widetilde{\n x}_p)$ is a prior based on a fixed design matrix; for example, $g$-Zellner type priors \citep{Zell:86, Zell:Siow:80, Baetal11, liang08, FLS01}, spike and slab priors \citep{IshwaRao05}, and non-local priors \citep{JohRos10}, among others.

However, for the case with missing data, it is customary to consider the covariates as random as it is explicitly assumed with \eqref{regres.gamma.2} (sometimes referred to as the imputation model). It plays a central role in dealing with missing data, and there is substantial literature on imputation models to accommodate different types of variables, as we mentioned in the introduction. Any of these can be used as desired without affecting the methods in this paper. We will use a multivariate normal imputation model in the section devoted to experiments.

Finally, to complete the probabilistic structure of the competing models, we must specify how the missing observations occur. The generally accepted framework that we adopt was introduced in \cite{LitRub20} and assumes that originally data are fully observed but some observations are hidden for the analyst. The process by which some observations are hidden is unknown, which leads us to consider $M$ as a random matrix whose modeling is specified in \eqref{regres.gamma.3} where $\n \psi$ are the parameters governing the missing mechanism.

For every model $\n\gamma\in\Gamma$---defined in \eqref{regres.gamma.1}--\eqref{regres.gamma.3}---and given a prior distribution for the unknown parameters $\pi_\gamma(\n\alpha,\n\beta_\gamma,\n\nu,\n\psi)$, we can now proceed to obtain the posterior probability for each model using Equation~\eqref{eq.post2} where
\begin{equation}\label{eq:margM}
	\begin{split}
		m_\gamma(&\widetilde{\n y}_{(0)},\widetilde{\n x}_{(0)}, \widetilde{M})=\\
		&\int f_\gamma(\widetilde{\n y}_{(0)}, {\n y}_{(1)}\mid \widetilde{\n x}_{(0)},{\n x}_{(1)}, \n\alpha,\n\beta_\gamma) f(\widetilde{\n x}_{(0)},{\n x}_{(1)}\mid\n\nu)  f(\widetilde M\mid \widetilde{\n y}_{(0)}, {\n y}_{(1)}, \widetilde{\n x}_{(0)}, {\n x}_{(1)},\n\psi)\\
		&\times \pi_\gamma(\n\alpha,\n\beta_\gamma,\n\nu,\n\psi)\mbox{d}[\n\alpha,\n\beta_\gamma,\n\nu,\n\psi,{\n x}_{(1)},{\n y}_{(1)}].
	\end{split}
\end{equation}

\subsection{Ignorability of the missing data mechanism}
We inspect the conditions under which Bayes factor, and hence the posterior probabilities of models, remains unaffected by the specific form of the missing data mechanism \eqref{regres.gamma.3}. 

According to \cite{VanB:18}, missing at random (MAR) holds if
\begin{equation}\label{MARat}
	f(\widetilde M\mid \widetilde{\n y}_{(0)}, \widetilde{\n x}_{(0)}, {\n y}_{(1)}, {\n x}_{(1)},\n\psi)=f(\widetilde M\mid \widetilde{\n y}_{(0)}, \widetilde{\n x}_{(0)},\n\psi),
\end{equation}
while missing completely at random (MCAR) holds if the distribution of $M$ does not depend on the observed data. 

In the next result we prove that, under the MAR assumption (or the more restrictive MCAR), the distribution of $M$ is ignorable. 

\begin{pro}\label{pro.igno}\mbox{ }\\
	If we assume MAR, and for all $\n\gamma\in\Gamma$
	the prior distribution satisfies:
	\begin{equation}\label{eq:priorgeneral}
		\pi_\gamma(\n\alpha,\n\beta_\gamma,\n\nu,\n\psi)=\pi_\gamma(\n\alpha,\n\beta_\gamma,\n\nu)\pi(\n\psi).
	\end{equation}
then
	\begin{equation}
	\label{eq.mgamma.all}
	m_\gamma(\widetilde{\n y}_{(0)},\widetilde{\n x}_{(0)}, \widetilde{M})\propto m_\gamma(\widetilde{\n y}_{(0)},\widetilde{\n x}_{(0)}),
	\end{equation}
	where
	\begin{equation}
	\label{eq.mgamma.igno}
	m_\gamma(\widetilde{\n y}_{(0)},\widetilde{\n x}_{(0)})=
	\int f_\gamma(\widetilde{\n y}_{(0)}\mid \widetilde{\n x}_{(0)},{\n x}_{(1)}, \n\alpha,\n\beta_\gamma)\,
	f(\widetilde{\n x}_{(0)},{\n x}_{(1)}\mid\n\nu) \pi_\gamma(\n\alpha,\n\beta_\gamma,\n\nu) \mbox{\normalfont d}[\n\alpha,\n\beta_\gamma,\n\nu,
	{\n x}_{(1)}].
	\end{equation}
\end{pro}
\begin{proof}
    Proofs can be found in the Supplementary material.
\end{proof}

Condition \eqref{eq:priorgeneral} requires that  the parameters in the regression and imputation components of the model are independent (a priori) of the parameters governing the missing data mechanism. In what follows in this paper, we assume MAR and that this condition holds. This combination exempts us from specifying \eqref{regres.gamma.3} and $\pi_\gamma(\n\psi)$ in virtue of Proposition~\ref{pro.igno}. We will revisit this hypothesis in the discussion in the concluding section. 


\subsection{A recognizable expression for marginals with missing data} \label{MURM.marginals}
Normally, procedures for dealing with missing data follow the logic of being extensions of a complete data method, with missing values replaced by some type of imputation. At first glance, the relevant marginal $m_\gamma(\n y_{(0)},\n x_{(0)})$ defined in \eqref{eq.mgamma.igno}, shows no evidence of this logic. Next, we derived an equivalent expression interpreted in this manner. 

\begin{res}\label{eq.margRR}
	Up to a proportionality constant common to all models, an equivalent expression for \eqref{eq.mgamma.igno} is 
	\begin{equation}
			m_\gamma(\n y_{(0)}, \n x_{(0)})\propto \int \mathfrak{m}_\gamma(\n y_{(0)} \mid \n x_{(0)},{\n x}_{(1)},\n\nu) \pi({\n x}_{(1)},\n\nu\mid \n x_{(0)})\, \mbox{\normalfont{d}}[{\n x}_{(1)},\n\nu] \label{marg.reg.gen}
	\end{equation}
	where $\pi({\n x}_{(1)}, \n\nu\mid \n x_{(0)})$ is the posterior distribution of $({\n x}_{(1)},\n\nu)$ given ${\n x}_{(0)}$, and
	\begin{equation}\label{cal.mar}
			\mathfrak{m}_\gamma(\n y_{(0)}\mid \n x_{(0)}, {\n x}_{(1)},\n\nu)=\int f_\gamma(\n y_{(0)} \mid \n x_{(0)},{\n x}_{(1)}, \n\alpha,\n\beta_\gamma)\pi_\gamma(\n\alpha,\n\beta_\gamma\mid\n\nu) \mbox{\normalfont{d}}[\n\alpha,\n\beta_\gamma].
	\end{equation}
	If model $\n\gamma$ has only common parameters, then the expression becomes:
	\begin{equation}\label{cal.mar0}
			\mathfrak{m}_\gamma(\n y_{(0)}\mid \n x_{(0)},{\n x}_{(1)},\n\nu)=\int f_\gamma(\n y_{(0)} \mid \n x_{(0)},{\n x}_{(1)}, \n\alpha) \pi_\gamma(\n\alpha\mid\n\nu) \mbox{\normalfont{d}}\n\alpha.   
	\end{equation}
	
\end{res} 
\begin{proof}
    Proofs can be found in the Supplementary material.
\end{proof}
Above, $\mathfrak{m}_\gamma$---a function of $\n\nu$ and $\n x_{(1)}$---is the ``missing data'' counterpart of the corresponding marginal used in the full data case for the calculation of BF (cf. Equation~\ref{mgamma.complete}). Thus, Equation~\eqref{cal.mar} states that $m_\gamma(\n y_{(0)}, \n x_{(0)})$ is the expected value of such ``missing data marginal'' with respect to the posterior distribution $\pi({\n x}_{(1)},\n\nu\mid \n x_{(0)})$ (which does not depend on $\n\gamma$ and that only involves observed covariates).

For a model $\n\gamma$, whose regression component \eqref{regres.gamma.1} is not affected by missingness---does not depend on ${\n x}_{(1)}$---it is clear that $\mathfrak{m}_\gamma(\n y_{(0)}\mid \n x_{(0)},{\n x}_{(1)},\n\nu)\equiv\mathfrak{m}_\gamma({\n y}_{(0)}\mid {\n x}_{(0)},\n\nu)$. Therefore, after integrating ${\n x}_{(1)}$ into Equation~\eqref{marg.reg.gen}, a simpler expression for $m_\gamma({\n y}_{(0)}, {\n x}_{(0)})$ can be derived as
\begin{equation}\label{partcase}
	m_\gamma({\n y}_{(0)}, {\n x}_{(0)})\propto \int \mathfrak{m}_\gamma({\n y}_{(0)}\mid {\n x}_{(0)},\n\nu)\,\pi_\gamma(\n\nu\mid {\n x}_{(0)})\, \mbox{d}\n\nu.
\end{equation}

\subsection{Computing the marginal by simulation} \label{marginal.approx.simul}

If a manageable expression for $\mathfrak{m}_\gamma({\n y}_{(0)}\mid \n x_{(0)},{\n x}_{(1)},\n\nu)$ is available, the marginal $m_\gamma({\n y}_{(0)}, {\n x}_{(0)})$ can be approximated with a Monte Carlo-based routine, as follows:

For $j=1,\ldots,J$: 
\begin{itemize}
	\item[] Step 1: Draw $\n\nu^{(j)}\sim\pi(\n\nu\mid {\n x}_{(0)})$,
	\item[] Step 2: Draw ${\n x}_{(1)}^{(j)} \sim f({\n x}_{(1)}\mid {\n x}_{(0)}, \n\nu^{(j)})$,
	\item[] Step 3: Calculate $\mathfrak{m}^{(j)}=\mathfrak{m}_\gamma({\n y}_{(0)}\mid {\n x}_{(0)},{\n x}_{(1)}^{(j)},\n\nu^{(j)})$,
\end{itemize} 
then compute $m_\gamma({\n y}_{(0)}, {\n x}_{(0)}) \approx J^{-1}\sum \mathfrak{m}^{(j)}$. The implementation of Step~2 can be approached with standard augmented Gibbs schemes  \citep[see, for instance,][]{hoff2009first}.

\subsection{Objective prior distributions on model parameters: general considerations}\label{Priors} 
The standard Bayesian method for addressing the absence of prior information uses improper distributions. In estimation problems (the model is fixed), the impropriety of priors does not imply any additional difficulty as long as the posterior is proper. There is a large body of literature regarding which priors are best suited to different models \citep[consult the catalogue][]{YangBer97}. Many of these can be obtained with mathematical rules \citep[like Jeffreys' priors or reference priors; see][]{ref:kass-wasserman-1996}. We refer to such (objective for estimation and usually improper) priors with the superindex $N$.

In the case of model uncertainty, the situation is quite different, and priors need to be carefully specified. In 
the Supplementary material
we discuss in depth about the structure of the prior, finally recommending:
\begin{equation}\label{my.prior.reg}
	\pi_\gamma(\n\alpha,\n\beta_\gamma,\n\nu)=\pi^N(\n\alpha)\pi^N(\n\nu)\pi_\gamma(\n\beta_\gamma\mid\n\nu,\n\alpha),
\end{equation}
reducing to
\begin{equation}\label{my.prior.common}
	\pi_\gamma(\n\alpha,\n\nu)=\pi^N(\n\alpha)
 \pi^N(\n\nu).
\end{equation}
for models with only common parameters. Above, the only ingredient that remains unspecified is $\pi_\gamma(\n\beta_\gamma\mid\n\nu,\n\alpha)$ which must be proper. We determine this distribution for the two problems considered in this paper in the following sections.


\section{Variable selection in the general linear model}\label{sec.lm}


\subsection{Model comparison}\label{sec:model.comp}
For $\n\gamma\ne \n 0$ let $X_{\gamma}$ be the sub-matrix of $X$ of dimension $n\times p_{\gamma}$ containing the $p_\gamma=\sum_{j=1}^p\gamma_j$ covariates corresponding to the ones in $\n\gamma$ . Let's consider the problem of selecting between two models of the form \eqref{regres.gamma.1}--\eqref{regres.gamma.3} where $\Gamma=\{\n 0,\n\gamma\}$ with
\begin{equation*}
	f_0(\n y \mid \n x_1,\ldots,\n x_p,\alpha,\sigma)=N_n(\n y\mid \alpha\n 1,\sigma^2 I),\;\;
	f_{\gamma}(\n y\mid \n x_1,\ldots,\n x_p,\alpha,\sigma,\n\beta)= N_n(\n y\mid \alpha\n 1+X_{\gamma}\n\beta_{\gamma},\sigma^2 I),
\end{equation*}

\begin{defn} We define the imputation $g'$-prior as
\begin{equation}\label{imput-gprior}
\pi_\gamma(\n\beta_\gamma\mid \alpha,\sigma,\n\nu)=N_{p_\gamma}(\n\beta_\gamma\mid \n 0, g'\,\sigma^2\,\Sigma_{\gamma\gamma}^{-1}),
\end{equation}
where $\Sigma_{\gamma\gamma}$ denotes the $p_{\gamma} \times p_{\gamma}$ block matrix from $\Sigma\equiv\Sigma(\n\nu)=V(\n x\mid\n\nu)$ corresponding to the active variables in $\n\gamma$.
\end{defn}
The arguments that lead to the definition of the imputation $g'$-prior are elaborated in Section~\ref{sub.gprimeconst}.

For the case with no-missing data the $g$-Zellner prior 
$
N_{p_\gamma}(\n\beta_\gamma\mid \n 0, g\, \sigma^2\,[(\overline{X}_{\gamma})^t \overline{X}_{\gamma}]^{-1})
$
with $g=n$ \citep[also called the unit information prior, see][]{Kass:Wass:95} can be interpreted as an empirical version of the imputation $g'$-prior with $g'=1$, as the covariance matrix in the $g$-prior (except for the $\sigma^2$) converges to $\Sigma_{\gamma\gamma}$ (if, as in the case with the unit information prior, $g\rightarrow\infty$). This limiting coincidence justifies our conventional choice $g'=1$ that we use in our numerical experiments. Alternatively, $g'$ can be seen as a hyperparameter---as in eg. \cite{Liang_2007}---with prior distribution $g'\sim \pi(g')$ leading to an hyper-$g'$ imputation prior. We leave the exploration of this further generalization for a future research.

\begin{pro}\label{prop.impgBF}
The  imputed $g'$-Bayes factor---corresponding to the prior under the scheme  \eqref{my.prior.reg}, \eqref{my.prior.common} and \eqref{imput-gprior}---, $B_{\gamma 0}(\n y_{(0)}, \n x_{(0)})$ is
\begin{equation}\label{eq:g-BF3}
E\Big\{
	\Big[\frac{S_0}{S_0-\n y_{(0)}^t \overline{X}_{\gamma}\big((\overline{X}_{\gamma})^t \overline{X}_{\gamma}+ \frac{1}{g'}\Sigma_{\gamma\gamma}\big)^{-1}{(\overline{X}_{\gamma})^t}\n y_{(0)}}\Big]^{(n_0-1)/2}
	\, \Big|{(\overline{X}_{\gamma})^t} \overline{X}_{\gamma}\;\Sigma_{\gamma \gamma}^{-1}+\frac{I}{g'} \Big|^{-1/2}
	\Big\},
\end{equation}
where the expectation is with respect to the posterior distribution $\pi^N(\n x_{(1)}, \n\nu\mid\n x_{(0)})$; $\overline{X}_{\gamma}=(I-\frac{1}{n_0}\n 1\n 1^t)X_{\gamma,n_0}$, and $X_{\gamma,n_0}$ is the matrix composed by the rows of $X_{\gamma}$ for which the dependent variable $y$ has been observed; $n_0$ is the length of $\n y_{(0)}$, $S_0$ is $n_0$ times the variance of $\n y_{(0)}$ and $I$ is the $p_\gamma \times p_\gamma$ identity matrix.
\end{pro}	

Again, if all data were observed, $B_{\gamma 0}(\n y_{(0)}, \n x_{(0)})$ generalizes the Bayes factor associated with the $g$-prior
\begin{equation}\label{completegBF}
\Big(1+n\frac{S_\gamma}{S_0}\Big)^{-(n-1)/2}\Big(1+n\Big)^{(n-p_\gamma-1)/2},
\end{equation}
(where $S_\gamma$ is the residual sum of squared errors for $\gamma$) which is obtained if we replace $\Sigma_{\gamma\gamma}$ by $n^{-1}(\overline{X}_{\gamma})^t \overline{X}_{\gamma}$ in \eqref{eq:g-BF3}.

\subsection{The construction of the imputation $g'$-prior}\label{sub.gprimeconst}




With respect to the common parameters $(\alpha,\sigma)$, as discussed in 
the Supplementary material, the use of the same prior in both models is reasonable if these parameters represent similar magnitudes in both models requiring a reparameterization in the model. 
In particular, we need to reparameterize the intercept to justify the assumption of a similar meaning. The idea is to transfer the mean of $\n x$ to the intercept such that it  has zero mean, as follows:
\begin{eqnarray*}
	\alpha\n 1+X_{\gamma}\n\beta_{\gamma}&=&\alpha\n 1+X_{\gamma}\n\beta_{\gamma}+\n\mu_{\gamma}^t\beta_{\gamma}\n 1-\n\mu_{\gamma}^t\beta_{\gamma}\n 1= (\alpha+\n\mu_{\gamma}^t\beta_{\gamma})\n 1+(X_{\gamma}-\n 1\n\mu_{\gamma}^t)\n\beta_{\gamma}=\\
	&\mathrel{\stackrel{\makebox[0pt]{\mbox{\normalfont\tiny def}}}{=}}&\alpha^\star\n 1+(X_{\gamma}-\n 1\n\mu_{\gamma}^t)\n\beta_{\gamma},
\end{eqnarray*}
where $\n\mu\equiv\n\mu(\n\nu)=E(\n x\mid\n\nu)$ is the mean of $\n x$ as obtained from the imputation model \eqref{regres.gamma.2}. 
With this reparameterization, the model $\n\gamma$ is redefined as:
\begin{equation}\label{M1.reparam}
		f_{\gamma}(\n y\mid \n x_1,\ldots,\n x_p,\alpha^\star,\n\beta_{\gamma},\sigma)= N_{n}(\n y\mid \alpha^\star\n 1+(X_{\gamma}-\n 1\n\mu_{\gamma}^t)\n\beta_{\gamma},\sigma^2 I).
\end{equation}
Now, the parameter $\alpha^\star$ (in $\n\gamma$) is the mean of $y$ when the values of the covariates coincide with their expectations, which aligns with the meaning of $\alpha$ in the null model (which represents the mean of $y$ regardless of the values of the covariates). This justifies using the same prior distribution (informative or objective) for $\alpha$ and $\alpha^\star$. Note that this result is achieved when the columns of $X_{\gamma}$ are centered with respect to their expectations, which is the counterpart to centering with respect to their sample means, as is routinely done in the literature. 

The above argument is rather informal but was used in the early literature on Bayesian testing, such as \cite{Jeff:61} or \cite{Zell:Siow:80}. More recently, \cite{Kass:Raft:95} worked on formalizing the concept of common parameters with similar meanings. They reasoned that such an assumption is sensible when the common and new parameters are orthogonal (i.e., the expected Fisher information matrix is block diagonal). In this case, the common parameters represent the same quantities, opening the possibility of using the same prior for both. When the covariates are random, the expected Fisher information matrix, $\mathfrak{I}$, for the parameters involved in the regression component of the model $\n\gamma$ (after the integration of $\n y$) is obtained over the imputation model of the covariates. In particular, 
\begin{equation}\label{FishIM}
	\mathfrak{I}= \frac{1}{\sigma^2}\, E\left(
	\begin{array}{ccc}
		n & \n 1^t(X_{\gamma}-\n 1\n\mu_{\gamma}^t) & 0\\
		(X_{\gamma}-\n 1\n\mu_{\gamma}^t)^t\n 1 & (X_{\gamma}-\n 1\n\mu_{\gamma}^t)^t (X_{\gamma}-\n 1\n\mu_{\gamma}^t) & \n 0\\
		0 & \n 0^t & 2n
	\end{array}\right)=\frac{n}{\sigma^2}\big(1 \oplus \Sigma_{\gamma\gamma} \oplus 2\big),  
\end{equation}
where $\Sigma_{\gamma\gamma}$ denotes the $p_{\gamma} \times p_{\gamma}$ block diagonal from $\Sigma\equiv\Sigma(\n\nu)=V(\n x\mid\n\nu)$ corresponding to the active variables in $\n\gamma$. We conclude that $\n\beta_{\gamma}$ and $(\alpha^\star,\sigma)$ are orthogonal, and that if $\pi_0(\alpha,\sigma)$ is used for the null model, we can use $\pi_{\gamma}(\alpha^\star,\sigma)=\pi_0(\alpha^\star,\sigma)$ for the alternative model. Note that this orthogonality does not hold for the original parameterization $(\alpha,\sigma)$. In the absence of prior information, the obvious choice in this case is the reference priors $\pi_0(\alpha,\sigma\mid\n\nu)=\sigma^{-1}$ and $\pi_{\gamma}(\alpha^\star,\sigma\mid\n\nu)=(\sigma)^{-1}$, which do not depend on the parameters of the distribution for the covariates, $\n \nu$. Our ultimate goal is to obtain the marginals where all parameters are integrated. Hence, 
$\star$ can be removed from the notation. What remains is the need to work with the alternative model in \eqref{M1.reparam}, in which the covariates are centered around their expected values. This must mimic the practice of centering the covariates around their sample means (which cannot be done with missing data).

Once we have established the prior for the common parameters, we now determine the prior $\pi_{\gamma}(\n\beta_{\gamma}\mid\alpha,\sigma,\n\nu)$. The extensive literature on $g$-priors agrees that we should use a $p_\gamma$-multivariate normal density (perhaps mixed to obtain flat tails) centered at zero and with a unitary covariance matrix $V$. This matrix is defined as the block corresponding to the inverse of the Fisher information matrix multiplied by sample size, $n$. For a complete dataset, this route leads to the use of $V=n\sigma^2\,(\overline{X}_{\gamma}^t\overline{X}_{\gamma})^{-1}$ (where $\overline{X}_{\gamma}$ has columns centered around the sample mean), as proposed in \cite{Zell:Siow:80} and unanimously followed in the related research \citep[see][and references therein]{Baetal11}. 

Mimicking this path in the case of missing data (or more in general for random covariates) is straightforward because we now have the expected Fisher information matrix. Furthermore, obtaining the inverse is rather simple because the matrix is block diagonal as a consequence of reparameterization (cf. Equation ~\ref{FishIM}), leading to 
$V=n\frac{\sigma^2}{n}\Sigma_{\gamma\gamma}^{-1}=\sigma^2\, \Sigma_{\gamma \gamma}^{-1}$. Remarkably, the sample size does not enter in the expression leading to \eqref{imput-gprior} with $g'=1$, as proposed.

\subsection{Variable Selection}
The basis for developing VS methods in the context of missing data is the two-model selection problem described in Section \ref{sec:model.comp}. In VS, the goal is to find which of the covariates $\{x_1,\ldots,x_p\}$ have a real effect on the response, $y$.

The list of possible models can be expressed using the binary parameter vector $\n\gamma^t=(\gamma_1,\ldots,\gamma_p)$, where $\gamma_j=1$ if the response depends on $x_j$ and zero otherwise. For example, a model with only $x_2$ corresponds to $\n\gamma^t=(0,1,0,\ldots,0)$. The set of possible models is denoted by $\Gamma$ and its cardinality is $2^p$, considering only the main effects. The posterior probability of each model  $\n \gamma$, as shown in Equation~\eqref{eq.post2}, depends on the prior probabilities over the model space. Some objective prior proposals are uniform, $p(\n \gamma)=1/2^p$ for $\n \gamma \in \Gamma$, or the hierarchical uniform prior discussed by \cite{Scott:Berg:10}: $p(\n \gamma)\propto 1/{p \choose p_\gamma},$ --recall  $p_{\gamma}=\sum_{j} \gamma_j$--. We strongly recommend the last prior because it accounts for the multiplicity of comparisons \citep{Scott:Berg:10}. 

The model posterior distribution $p(\n \gamma \mid \n y_{(0)}, \n x_{(0)})$ is the main tool for quantifying uncertainty in the VS problem and must be properly summarised to produce useful reports. Rather than selecting a single model, as in the case of model comparison, the posterior distribution offers an enormous variety of ways to gain insight into the primary question of measuring the effect of different covariates on the response. Common summaries are the highest probability model and its probability; the posterior inclusion probability of each individual variable, which for the $j$th covariate is 
$p(\gamma_j=1\mid \n y_{(0)}, \n x_{(0)})=\sum_{\gamma \in \Gamma: \gamma_j=1} p( \n \gamma \mid \n y_{(0)}, \n x_{(0)}),$ 
and the median probability model, which includes covariates with inclusion probabilities larger than $0.5$ \citep{Barbieri-etal21, Barb:Berg:04}. 

Finally, the posterior distribution provides straightforward access to (Bayesian) Model Averaged estimations and predictions as described in \cite{Hetal99} or \cite{Steel20}.

\section{Uncertainty on the distributions of the errors}\label{sec.errors}
\paragraph{Model comparison.}
Let $X_1$ be an $n\times p_1$ matrix containing certain subset of the covariates in $X$ (possibly with missing cells). We consider the problem where competing models agree on the covariates but differ in the density assumed for the errors.
Consequently, we have two candidate models of the form \eqref{regres.gamma.1}--\eqref{regres.gamma.3}
where
$$
f_\gamma(\n y\mid \n x_1,\ldots,\n x_p,\alpha,\sigma,\n\beta_1)=\sigma^{-n}\, h_\gamma\big(\frac{\n y-\n 1\alpha-X_1\n\beta_1}{\sigma}\big),\;\; \gamma\in\Gamma=\{1,2\}
$$
and $h_1,h_2:\erre^n\rightarrow \erre$ are known (multivariate) probability density functions. In this problem, there is uncertainty regarding the distribution of the errors (e.g., a multivariate normal versus a multivariate Student's t or, as in the accompanying example, testing a particular heteroscedastic form).
Following the generic notation in Section~\ref{MURM} the common parameters are $\n\alpha\equiv(\alpha,\sigma,\n\beta_1)$ and $\n\nu$ whereas there are no new parameters.

Following the arguments below, the priors we propose are
\begin{equation}\label{priors.errors}
\pi_\gamma(\alpha,\n\beta_1,\sigma,\n\nu)=\pi^N(\alpha,\n\beta_1,\sigma)\pi^N(\n\nu)=\sigma^{-1}\pi^N(\n\nu).
\end{equation} 
From these, the imputation Bayes factor is the ratio of marginals
\begin{equation}\label{BF.errors}
B_{12}(\n y_{(0)}, \n x_{(0)})=\frac{E\{\mathfrak{m}_1(\n y_{(0)}\mid \n x_{(0)}, \n x_{(1)}, \n\nu)\}}{E\{\mathfrak{m}_2(\n y_{(0)}\mid \n x_{(0)}, \n x_{(1)}, \n\nu)\}},
\end{equation}
where both expectations are with respect to $\pi^N(\n x_{(1)}, \n\nu\mid\n x_{(0)})$ and
\begin{equation*}
	\mathfrak{m}_\gamma(\n y_{(0)}\mid \n x_{(0)}, \n x_{(1)}, \n\nu)=  \int \sigma^{-n_0}\, h_\gamma\left(\frac{\n y_{(0)}-\n 1\alpha-X_1\n\beta_1}{\sigma}\right)\frac{1}{\sigma}\, \mbox{d}[\alpha, \n\beta_1,\sigma], 
\end{equation*}
for $\gamma=1,2$. Notice that, in this case, $\mathfrak{m}_\gamma$ does not depend on $\n\nu$---only on $\n x_{(1)}$---and hence the expectation in \eqref{BF.errors} is with respect to the \emph{a posteriori} predictive distribution $\pi^N(\n x_{(1)}\mid\n x_{(0)})$.

\emph{The construction of the imputation prior.} First notice that both competing models share a common group of invariance. More concisely, they are group-invariant with respect to transformations of type \citep[see for example, ][]{Eato:1989}:
$
\{\n y\rightarrow c\n y+[\n 1\, X_1]\, \n b,\,\, c\in\erre; \n b\in\erre^m\}.
$
This property has two main consequences on the priors that we will highlight in the next paragraph.

There are only common parameters in this problem, so the starting point is the recommended scheme \eqref{my.prior.common}, which in our problem leads to \eqref{priors.errors},
where $\sigma^{-1}$ is chosen because it is the right Haar measure associated with the said type of invariance (first consequence). Additionally, the justification for using the \emph{same} prior for $\alpha,\n\beta_1,\sigma$ relies on the argument that these parameters have the same dimension and common meaning regarding their roles within the aforementioned shared invariance structure (second consequence). For instance, $\sigma$ acts as a scale parameter in both models, whereas $\alpha$ is the location parameter. Remarkably, this informal reasoning was supported by formal arguments from \cite{Berg:Peri:Vars:98}, who perhaps provided one of the most important results for objective priors within model uncertainty. These authors showed that under very soft conditions on $h_\gamma$ and under the conditions of ``shared invariance'' mentioned above, the right Haar density provides an exact predictive match \citep[see also][]{Baetal11}.

\begin{example} \label{example.errors}
	In this example, we test for possible he\-te\-roscedasticity in the errors comparing
	\begin{equation}\label{distrib.errors}
	h_1(\n\varepsilon)=N_n(\n\varepsilon\mid \n 0, I),\,\, h_2(\n\varepsilon)=N_n(\n\varepsilon\mid 0,\Psi),
	\end{equation}
	where $\Psi$ is a known positive definite matrix. If there are no missing values for the dependent variable, it is straightforward to derive that $\mathfrak{m}_\gamma(\n y\mid \n x_{(0)}, \n x_{(1)}, \n\nu)$ has a closed-form expression leading to
	\begin{equation}\label{eq:g-BF2}
B_{12}(\n y, \n x_{(0)})=
\frac{E\{|(1\, X_1)^t(1\, X_1)|^{-1/2}\}}{E\{\frac{1}{S_\Psi^{(n-p_1-1)/2}}
	\frac{|(1\, X_1)^t\Psi^{-1}(1\, X_1)|^{-1/2}}{|\Psi|^{-1/2}}\}},	
	\end{equation}
	where
	\begin{equation*}
			S_\Psi=\n y^t\Big(\Psi^{-1}-\Psi^{-1}(1X_1)\big((1X_1)^t\Psi^{-1}(1X_1)\big)^{-1}(1X_1)^t\Psi^{-1}\Big)\n y
	\end{equation*}
	(sum of the squared errors when regressing $\Psi^{-1/2}\n y$ with the columns $\Psi^{-1/2}(1\, X_1)$). 
	
	 If $\n y$ had missing observations, the expressions would be similar, replacing $\n y$ with $\n y_{(0)}$ and $n$ with $n_{0}$ and selecting the rows corresponding to the observed units in $(1\, X_1)$ and $\Psi$.
	
\end{example}

\section{Numerical experiments} \label{sec.example}
We conducted several experiments to shed light on the implications of missing observations in model uncertainty problems. This study attempts to fill a gap in the literature where the evidence thus far is limited, especially from a Bayesian perspective. We performed five experiments based on the general linear model but of quite a different nature, ranging from highly controlled simulated cases to real datasets. The first four experiments considered the uncertainty of the regressors, while the fifth experiment questioned the structure of the error covariance. For comparisons, in all cases we have access to the full dataset (before missingness occurs).

In all cases, we use a multivariate normal imputation model. That is, \eqref{regres.gamma.2} is
$$
(x_{i1},x_{i2},\ldots,x_{ip})\mid \n\nu \sim N_p(\n\mu,\Sigma) 
$$
where $\n\nu=(\n\mu,\Sigma)$. Some of our experiments are based on real data with covariates far from being normal (see Experiment S2 of Supplementary material), hence allowing to analyze the effect of a bad imputation model in the posterior distribution. The reference prior that corresponds to the the multivariate normal distribution is derived in \cite{ChEa90}:
$\pi^N(\n\mu,\Sigma)=|\Sigma|^{-(p+1)/2}|I+\Sigma*\Sigma^{-1}|^{-1/2}$, 
 where $*$ denotes the Hadamard product (component by component). The corresponding posterior distribution has no closed form, but it can be sampled easily using the simple rejection algorithm described in \cite{SunBer06}.


Experiments 1, 2, S1 and S2 concern Section~\ref{sec.lm} and we refer to the \emph{oracle $g$-BF} to the Bayes factor \eqref{completegBF}---corresponding to the $g$-prior with $g=n$---using the full dataset. Similarly, the same Bayes factor applied to the dataset resulting after listwise deletion is termed as \emph{listwise deletion $g$-BF}. Finally, our proposed Bayes factor, which utilizes all the available data by means of \eqref{eq:g-BF3} is the \emph{imputed $g'$-Bayes factor}.


The corresponding software can be found as a shiny application\footnote{\url{https://stefanocabras.shinyapps.io/muqmissing/}}, and the core code is available on github\footnote{\url{https://github.com/scabras/muqmissing}} along with the other pieces of code mentioned below.

\subsection{Experiment 1. Variable selection}
In this section, we reproduce the simulated experiment of \cite{Biometrics2005} to compare our results with SIAS (see Section~\ref{sec.RR}), which showed the best performance among the methods compared in that paper. For a comprehensive comparative study, we added the results for the full dataset (referred to as the oracle) and listwise deletion. 

The experiment consisted of $p=10$ potential explanatory variables, $x_1, \ldots, x_{10}$, simulated independently of a multivariate normal, where the off-diagonal elements of the correlation matrix were $\rho\in\{0.1,0.5\}$ (defining two different scenarios). This is combined with two ignorable missing data mechanisms: the MCAR mechanism, where values are randomly dropped from $x_j$, $j=1,\ldots, 10$ independently with a probability of either 5\% or 10\%, resulting in a global missing percentage (i.e., the proportion of individuals with at least one missing value in any covariate) of 40\% and 65\%, respectively, and an MAR, where $x_1, \ldots, x_5$ are fully observed, while amputation is performed over $x_j$, $j=6,\ldots,10$, with the same overall percentages of missing data as before, that is, 40\% and 65\%. For the latter scenario, we use the {\tt ampute} function from the {\tt mice} package in R, considering different missing patterns with 20\% or 40\% missing data for each variable to obtain the desired global missing percentages. 

The response variable $y$ was simulated using the following linear regression model:
$$
y_i = x_{i1} + 2 x_{i2} + x_{i6} + 2x_{i7} + \varepsilon_i, \quad \varepsilon_i\sim N(0,\sigma^2=2.5).
$$
For each combination of $\rho\in\{0.1, 0.5\}$, the overall percentage of missingness (40\%, 65\%), and missing data mechanism (MCAR, MAR), 100 datasets were simulated.

We calculated the posterior inclusion probabilities based on the oracle $g$-BF, listwise deletion $g$-BF, and imputed $g'$-BF, to which we appended the results reported in \cite{Biometrics2005} corresponding to the SIAS method. Following \cite{Biometrics2005}, we also compute a summary statistic, the signal-to-noise ratio (SNR), to compare the discriminatory power of the procedures, namely the ratio of the minimum inclusion probability for true covariates to the maximum for spurious predictors:
$$
\mbox{SNR} = \frac{\underset{j \in \{1,2,6,7\}}{\mbox{ min }}p(\gamma_j=1 \mid \n y_{(0)}, \n x_{(0)})}{\underset{j \in \{3,4,5,8,9,10\}}{\mbox{ max }}p(\gamma_j=1 \mid \n y_{(0)}, \n x_{(0)})}.
$$
Table~\ref{T1-sim-var-sel} shows the mean and standard deviation of each active variable's posterior inclusion probability and SNR for each combination of design elements. The conclusions drawn from this table are as follows:

\begin{table*}
	\caption{Experiment 1. Mean posterior inclusion probabilities for the truly active variables and mean signal-to-noise ratio. The number in parentheses corresponds to the standard deviation, reported only when $\geq 0.01$. Values for the SIAS method are borrowed from \cite{Biometrics2005}.}
	\label{T1-sim-var-sel}
	\centering
{\small\scalebox{0.8}{						
		\begin{tabular}{lcccccccccc} 
			&\multicolumn{5}{c}{$\rho=0.1$} & \multicolumn{5}{c}{$\rho=0.5$}\\
			& $x_1$ & $x_2$ & $x_6$ & $x_7$ & SNR & $x_1$ & $x_2$ & $x_6$ & $x_7$ & SNR \\ \hline 
			oracle & 1 & 1 & 1 & 1 & 3.8(.2) & 1 & 1 & 1 & 1 & 4.8(.2) \\
			\hline
			\multicolumn{11}{l}{40\%-MCAR}\\
			\hline
			Imputed & 1 & 1 & 1 & 1 & 3.5(.2) & 1 & 1 & 1 & 1 & 4.2(.2)\\
			Deletion & 1 & 1 & 1 & 1 & 3.3(.1) &.99(.05) & 1 &.99(.06) & 1 & 3.7(.2)\\
			SIAS &.92 &.99 &.91 &.99 & 2.9 &.80 &.99 &.80 &.99 & 2.4\\
			\hline
			\multicolumn{11}{l}{65\%-MCAR}\\
			\hline
			Imputed & 1(.01) & 1 & 1(.02) & 1 & 3.4(.2) & 1 & 1 & 1 & 1 & 3.9(.2)\\
			Deletion &.98(.06) & 1 &.95(.13) &1 & 2.6(.1) &.86(.18) & 1&.88(.18) &1(.02) & 2.6(.1)\\
			SIAS &.88 &.99 &.88 &.99 & 3.0 &.71 &.99 &.72 &.99 & 2.3\\
			\hline
			\multicolumn{11}{l}{40\%-MAR}\\
			\hline
			Imputed & 1(.02) & 1 & 1(.01) & 1 & 3.5(.2) & 1(.01) & 1 & 1(.01) & 1 & 3.9(.2)\\
			Deletion & 1 & 1 & 1(.02)& 1 & 3.0(.1) &.99(.05) & 1 &.99(.04) & 1 & 3.7(.2)\\
			SIAS &.90 &.99 &.90 &.99 & 2.8 &.88 &.99 &.77 &.99 & 2.1\\
			\hline
			\multicolumn{11}{l}{65\%-MAR}\\
			\hline
			Imputed &.99(.07) & 1 &.93(.14) & 1 & 3.2(.2) &.99(.03) & 1 &.92(.16) & 1 & 3.1(.2)\\
			Deletion &.97(.08) & 1 &.97(.08) & 1 & 2.6(.2) &.87(.18) & 1 &.88(.17) & 1 & 2.5(.1)\\
			SIAS &.82 &.99 &.85 &.99 & 2.4 &.89 &.99 &.69 &.98 & 1.5\\ \hline				
		\end{tabular}
}}
\end{table*}

The first conclusion is that listwise deletion performs competently and clearly outperforms SIAS. This is a surprising result, especially considering that this superiority occurs in all cases, both in the ability to preserve the strength of the true signals and in the discriminatory power (as measured by SNR). Furthermore, the differences between the two approaches are generally substantial. When comparing imputation and deletion, when the correlation between the covariates is small ($\rho=0.1$), the two approaches behave similarly in terms of sensitivity (the ability to detect true positives). As $\rho$ and the percentage of missingness increase, imputation outperforms listwise deletion, justifying the extra effort required in the procedure.

The imputation SNR was considerably better than the other methods (persistent in all cases and quite pronounced in some cases). Obviously, this is essentially a better performance in terms of specificity (the ability to detect true negatives) simply because the inclusion probabilities of signals are very close to one in the vast majority of cases. Compared with listwise deletion, this is also explained by the differences in the amount of sampling information used by each method. For example, consider the MCAR case with 65$\%$ missing data. Out of the $n\times (p+1)=1100$ total observations used by the oracle, listwise deletion preserves $0.35\times 1100=385$, whereas the results based on imputation use $1100-0.1\times 1000=1000$ (as the response variable is not imputed). This corresponds to approximately 2.6 times more sampling information, which, when accompanied by reliable imputations, leads to a substantial increase in specificity and sensitivity simply because the sample size is much larger.

Experiment S1 of the Supplementary material, although considering a simpler design, aimed to analyze the performance of our method by confronting it with listwise deletion in a more extreme case of the signal-to-noise ratio. 

\subsection{Experiment 2. The Ozone dataset}
We consider VS problems from popular real-world datasets in this and in Experiments S2 and S3 of the Supplementary material. The role of the distribution of the covariates, which is unknown in this case, is the main difference from previous simulated experiments. As a reminder, we assume a multivariate imputation model. Clearly, misspecification of this component does not affect listwise deletion procedures, but it is an essential part of all imputation methods. This observation is important for understanding the following results.

The Ozone datasets previously used in \cite{Ga-DoMa-Be13}, \cite{CasMor06}, and \cite{BerMol05} consisted of $n = 178$ measurements of atmospheric ozone concentration, along with several covariates. From the original 10 main effects, we only used seven with atmospheric relevance, which corresponds to the main effects in the {\tt Ozone35} dataset from the {\tt BayesVarSel} library in R, named $x_4$ to $x_{10}$. An initial examination of the data suggests that the assumption of normality is reasonable. For further details on these data, see \cite{CasMor06}. 

\begin{figure}
	\begin{center}
			\includegraphics[scale=0.7]{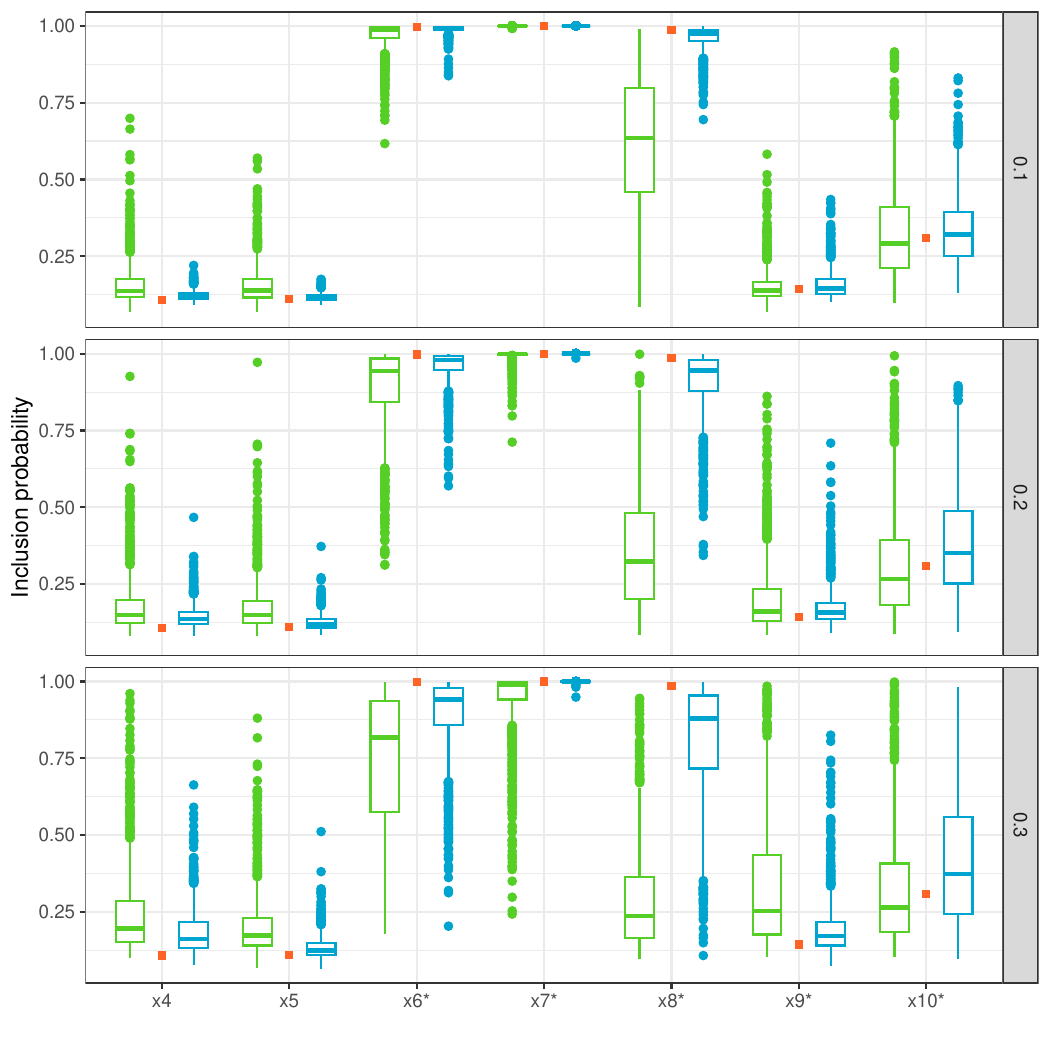} 
	\end{center}
		\caption{Boxplots of the inclusion probabilities for each variable using imputed $g'$-BF (blue) and listwise deletion $g$-BF (green), when considering 10 (top), 20 (middle) or 30\% (bottom) of missing values per variable, for Ozone dataset. The corresponding oracle $g$-BF inclusion probabilities are depicted in red. The symbol $\star$ in variable names explicitly indicates the variables with missing data.} 
	\label{all.bp}
\end{figure}

We introduced MAR NAs into variables $x_6$ to $x_{10}$ by using the function {\tt ampute} from the {\tt mice} package in R. The percentage of missing values per variable was 10, 20, or 30\%, resulting in a mean overall percentage of missingness of approximately 37\%, 60\%, and 74\%, respectively. For each of these percentages, we considered 1000 replications where the variability was caused by the removed observations (that changed in the replicas). Figure~\ref{all.bp} shows the variation in the inclusion probabilities for each variable obtained with the different Bayes factors.

The potential of the imputed $g'$-BF to preserve the evidence is shown in Figure~\ref{all.bp}. Its superiority over deletion was evident for all variables and levels of missingness. We also observed that the imputed $g'$-BF was less sensitive to variations in the removed observations, producing less variable results. 

Experiment S2 of the Supplementary material uses the Boston dataset, which includes variables that do not follow the normality assumption. This experiment helped us assess the performance of the proposed method under more challenging conditions where some of the assumptions may not hold. Finally, Experiment S3 of the Supplementary material focused on comparing the distributions of errors to illustrate the methods discussed in Section~\ref{sec.errors}.

\section{Conclusions and Future Work} \label{sec:conclusions}

This study presents a comprehensive approach for addressing model uncertainty when dealing with missing data in a regression framework. Through a series of experiments, we demonstrated the effectiveness of our proposed imputed $g'$-prior methodology compared to listwise deletion and the SIAS proposal from \cite{Biometrics2005}, particularly regarding reduced variability and more accurate posterior inclusion probabilities. The proposed method is fully automatic and does not depend on hyperparameters such as the penalty parameter of lasso methods. Moreover, exploiting the analytical integration from the closed-form output of the ``completed'' predictive distribution makes the method much faster and more efficient than the alternative procedure of imputing and estimating the model's posterior probability. There are several directions for future research to further enhance the applicability and robustness of our approach.

\begin{enumerate}
	
	\item \textit{Large model spaces}: Our experiments were conducted in relatively small model spaces, which allowed for exhaustive enumeration. However, in many modern applications, large model spaces (i.e., large $p$) are common, such as those arising from variable selection with many covariates. Adapting our approach to handle these situations would require the development of numerical algorithms, such as ``missing data" adaptations of Gibbs sampling methods \citep{Ga-DoMa-Be13}.
	
	\item \textit{Prior distribution for regression parameters}: A central question in this study has been the construction of objective prior distributions when the covariates are assumed to be random. Although arguably more realistic than the fixed design assumption, this perspective has barely received any attention in the literature despite the broad potential interest in such inferential objects. In this study, we followed \cite{Zell:86, Zell:Siow:80}, in what has been called $g$-priors, constructed based on the expected Fisher information matrix. We have derived a new class of $g$-priors in which we have focused on fixed $g$, but extensions to random hyperparameter \citep[of the type in][]{liang08} are straightforward. The construction of priors following alternative procedures, assuming that the covariates are random, opens up areas for future research that would extend the scope of non-local priors \citep{JohRos10}; modern spike and slab formulations \citep{Baietal21}; intrinsic priors \citep{Berg:Peri:96b,More:Bert:Racu:98} or power expected posterior priors \citep{FouNtz22}, to mention a few.

	\item \textit{Other patterns of missingness}: Our procedure can incorporate other missing data patterns, making it possible, in principle, to test for different missing data mechanisms. Further work in this direction is needed to understand how to separate the comparison of the missing data mechanism from the models for observed variables, response and covariates. 
	
	
\end{enumerate}

\newpage

\appendix{SUPPLEMENTARY MATERIAL}

\section{Proofs}

\subsection{Proof of Proposition~\ref{pro.igno}}
To prove Equation~\eqref{eq.mgamma.all}, we consider the definition of the predictive distribution in Equation~\eqref{eq:margM}.
\begin{eqnarray*}
	&&
	m_\gamma(\widetilde{\n y}_{(0)},\widetilde{\n x}_{(0)}, \widetilde{M})=\\
	&&\int f_\gamma(\widetilde{\n y}_{(0)}, {\n y}_{(1)}\mid \widetilde{\n x}_{(0)},{\n x}_{(1)}, \n\alpha,\n\beta_\gamma)\,
	f(\widetilde{\n x}_{(0)},{\n x}_{(1)}\mid\n\nu)f(\widetilde M\mid \widetilde{\n y}_{(0)}, {\n y}_{(1)}, \widetilde{\n x}_{(0)}, {\n x}_{(1)},\n\psi)\\
	&& \times \,\pi_\gamma(\n\alpha,\n\beta_\gamma,\n\nu,\n\psi)\mbox{d}[\n\alpha,\n\beta_\gamma,\n\nu,\n\psi,{\n x}_{(1)},{\n y}_{(1)}].
\end{eqnarray*}
As we assume MAR and consider that the prior distributions verify \eqref{eq:priorgeneral},
integrating $\n y_{(1)}$ (as $\n y_{(1)}$ and $\n y_{(0)}$ are conditionally independent given the covariates and the regression parameters), we obtain a simplified expression for $m_\gamma(\widetilde{\n y}_{(0)},\widetilde{\n x}_{(0)}, \widetilde{M})$:
\begin{eqnarray}\label{eq:proofprop1}
	&&m_\gamma(\widetilde{\n y}_{(0)},\widetilde{\n x}_{(0)}, \widetilde{M})=\nonumber\\
	&=&\int f_\gamma(\widetilde{\n y}_{(0)}\mid \widetilde{\n x}_{(0)}, \n x_{(1)}, \n\alpha,\n\beta_\gamma)\,
	f(\widetilde{\n x}_{(0)},{\n x}_{(1)}\mid\n\nu)\nonumber\\
	&\times&f(\widetilde M\mid \widetilde{\n y}_{(0)}, \widetilde{\n x}_{(0)},\n\psi) \pi_\gamma(\n\alpha,\n\beta_\gamma,\n\nu)\pi(\n\psi)\mbox{d}[\n\alpha,\n\beta_\gamma,\n\nu,\n\psi,{\n x}_{(1)}]\nonumber\\
	&=& \int f(\widetilde M\mid \widetilde{\n y}_{(0)}, \widetilde{\n x}_{(0)},\n\psi)\pi(\n\psi)\mbox{d}\n\psi \times m_\gamma(\widetilde{\n y}_{(0)},\widetilde{\n x}_{(0)}) \propto m_\gamma(\widetilde{\n y}_{(0)},\widetilde{\n x}_{(0)}), 
\end{eqnarray}
where
\begin{eqnarray*}
	&&m_\gamma(\widetilde{\n y}_{(0)},\widetilde{\n x}_{(0)})\nonumber\\
	&=&\int f_\gamma(\widetilde{\n y}_{(0)}\mid \widetilde{\n x}_{(0)},{\n x}_{(1)}, \n\alpha,\n\beta_\gamma)\,
	f(\widetilde{\n x}_{(0)},{\n x}_{(1)}\mid\n\nu) \pi_\gamma(\n\alpha,\n\beta_\gamma,\n\nu) \mbox{d}[\n\alpha,\n\beta_\gamma,\n\nu,
	{\n x}_{(1)}].
\end{eqnarray*}

Note that in Equation~\eqref{eq:proofprop1}, the first factor does not depend on $\gamma$ and cancels the Bayes factors and posterior probabilities of the models.

\subsection{Proof of Result~\ref{eq.margRR}}

Equation~\eqref{eq.mgamma.igno} can be expressed as 
\begin{eqnarray}
	&&m_\gamma(\n y_{(0)}, \n x_{(0)}) =\nonumber \\
	&=&\underbrace{\int \Big[f_\gamma(\n y_{(0)}\mid \n x_{(0)},{\n x}_{(1)}, \n\alpha,\n\beta_\gamma)\,\pi_\gamma(\n\alpha,\n\beta_\gamma\mid\n\nu)\, \mbox{d}[\n\alpha,\n\beta_\gamma] \Big]}_{\mathfrak{m}_\gamma(\n y_{(0)}\mid \n x_{(0)},{\n x}_{(1)},\n\nu)} \nonumber\\
	&&\times f(\n x_{(0)},{\n x}_{(1)}\mid\n\nu) \pi_\gamma(\n\nu) \, \mbox{d}[\n\nu, {\n x}_{(1)}] \nonumber\\
	&=& \int \mathfrak{m}_\gamma(\n y_{(0)} \mid {\n x}_{(0)},{\n x}_{(1)},\n\nu)\,f(\n x_{(0)},{\n x}_{(1)}\mid \n\nu)\pi_\gamma(\n\nu)\, \mbox{d}[\n\nu, {\n x}_{(1)}]\nonumber\\
	&=& \int \mathfrak{m}_\gamma(\n y_{(0)} \mid \n x_{(0)},{\n x}_{(1)},\n\nu)\,f({\n x}_{(1)}\mid \n x_{(0)}, \n\nu)f(\n x_{(0)}\mid \n\nu)\pi_\gamma(\n\nu)\, \mbox{d}[\n\nu, {\n x}_{(1)}]\nonumber\\
	&=& m_\gamma(\n x_{(0)})\, \int \mathfrak{m}_\gamma(\n y_{(0)} \mid \n x_{(0)},{\n x}_{(1)},\n\nu)\,f({\n x}_{(1)}\mid \n x_{(0)}, \n\nu)\pi_\gamma(\n\nu\mid \n x_{(0)})\, \mbox{d}[\n\nu, {\n x}_{(1)}] 
\end{eqnarray}
where $m_\gamma(\n x_{(0)})=\int f(\n x_{(0)}\mid \n\nu)\pi_\gamma(\n\nu)\, \mbox{d}\n\nu$ and $\pi_\gamma(\n\nu\mid \n x_{(0)})$ is the posterior distribution of $\n\nu$ given ${\n x}_{(0)}$.

\subsection{Proof of Proposition \ref{prop.impgBF}}

The marginals for the null model and $\n\gamma$ can be obtained using \eqref{marg.reg.gen}:
\begin{equation} \label{marg.lm}
	m_\gamma(\n y_{(0)}, \n x_{(0)})\propto
	\int \mathfrak{m}_\gamma(\n y_{(0)}\mid \n x_{(0)}, \n x_{(1)}, \n\nu)\, f(\n x_{(1)}\mid \n x_{(0)},\n\nu) \pi(\n\nu\mid\n x_{(0)})\, \mbox{d}[\n x_{(1)}, \n\nu]
\end{equation}
where
\begin{equation*}
	f(\n x_{(1)}\mid \n x_{(0)},\n\nu)
	\pi(\n\nu\mid\n x_{(0)})\propto \prod_{i=1}^n\, f((x_{i1}^I,x_{i2}^I,\ldots,x_{ip}^I) \mid\n\nu)\times \pi^N(\n\nu)
\end{equation*}
and $x_{ij}^I$ is the completed value (either imputed or originally observed) of individual $i$ and variable $j$. 
The null model depends only on the common parameters; therefore, using \eqref{cal.mar0}, 
\begin{equation*}
	\mathfrak{m}_0(\n y_{(0)}\mid \n x_{(0)}, \n x_{(1)}, \n\nu)=\int N_{n_0}(\n y_{(0)}\mid \alpha\n 1,\sigma^2 I)\frac{1}{\sigma}\, \mbox{d}[\alpha, \sigma].  
\end{equation*}

Finally, using \eqref{cal.mar},
\begin{equation*}
	\begin{split}
		\mathfrak{m}_{\gamma}(&\n y_{(0)}\mid \n x_{(0)}, \n x_{(1)}, \n\nu)=\\
		&\int N_{n_0}(\n y_{(0)}\mid \alpha \n 1+(X_{\gamma,n_0}-\n 1\n\mu_{\gamma}^t)\n\beta_{\gamma},\sigma^2 I) \frac{1}{\sigma}N_{p_{\gamma}}(\n\beta_{\gamma}\mid 0,g\sigma^2\Sigma_{\gamma \gamma}^{-1})\, \mbox{d}[\alpha,\n\beta_{\gamma},\sigma].
	\end{split}
\end{equation*}

In this problem the BF is
\begin{eqnarray*}
	B_{10}(\n y_{(0)}, \n x_{(0)})&=&\frac{m_1(\n y_{(0)}, \n x_{(0)})}{m_0(\n y_{(0)}, \n x_{(0)})}=
	\frac{E\big(\mathfrak{m}_1(\n y_{(0)}\mid \n x_{(0)}, \n x_{(1)}, \n\nu)\big)}{E\big(\mathfrak{m}_0(\n y_{(0)}\mid \n x_{(0)}, \n x_{(1)}, \n\nu)\big)}\\
	&=&
	E\Big(\frac{\mathfrak{m}_1(\n y_{(0)}\mid \n x_{(0)}, \n x_{(1)}, \n\nu)}{\mathfrak{m}_0(\n y_{(0)}\mid \n x_{(0)}, \n x_{(1)}, \n\nu)}\Big),
\end{eqnarray*}
where expectations are with respect to the posterior $\n x_{(1)}, \n\nu\mid\n x_{(0)}$. The last identity holds because $\mathfrak{m}_0$ is a constant for the expectation. Furthermore, the expression inside the large brackets is clearly reminiscent of an \emph{imputed} BF, such that $B_{10}(\n y_{(0)}, \n x_{(0)})$ is essentially the average of an imputed conventional BF (suggesting, for instance, a possible shortcut for its approximate computation using imputation and BF software in tandem).

Note that such an interpretation is appropriate because the null model does not depend on unobserved covariates. This is a usual situation, particularly in VS; however, it is not generally true. For example, a different situation occurs when  the null model depends on covariates with missing values.

The ratio $\mathfrak{m}_1/\mathfrak{m}_0$ has a closed-form expression as follows:
\begin{eqnarray}\label{eq:g-BF1}
	\frac{\mathfrak{m}_1(\n y_{(0)}\mid \n x_{(0)}, \n x_{(1)}, \n\nu)}{\mathfrak{m}_0(\n y_{(0)}\mid \n x_{(0)}, \n x_{(1)}, \n\nu)}&= &
	\Big[\frac{S_0}{S_0-\n y_{(0)}^t \overline{X}_{\gamma}\big((\overline{X}_{\gamma})^t \overline{X}_{\gamma}+ \Sigma_{\gamma\gamma}/g\big)^{-1}{(\overline{X}_{\gamma})^t}\n y_{(0)}}\Big]^{(n_0-1)/2} \nonumber  \\
	&\times& \Big|{(\overline{X}_{\gamma})^t} \overline{X}_{\gamma}\;\Sigma_{\gamma \gamma}^{-1}+1/g\, I\Big|^{-1/2},  \label{ratio.predictive.ex1}
\end{eqnarray}
where $\overline{X}_{\gamma}=(I-\frac{1}{n_0}\n 1\n 1^t)X_{\gamma,n_0}$, $S_0$ is $n_0$ times the variance of $\n y_{(0)}$, and $I$ is the $p_\gamma \times p_\gamma$ identity matrix.
Here, $X_{\gamma,n_0}$ denotes the matrix composed of the columns in model $\gamma$ and the rows for which $y$ has been observed; that is, the individuals at $y_{(0)}$. Note that although only $X_{\gamma}$ appears in the above expression, where missing values have been completed with the imputed values, the entire matrix $X$ with all its rows ($n$) is required to calculate the matrix $\Sigma_{\gamma\gamma}$. In par\-ti\-cu\-lar, all covariates, including the $p-p_{\gamma}$ columns of $X$, enter the process of imputing the missing values in $X_{\gamma}$ and obtain the posterior distribution $\n\nu \mid \n x_{(0)}$.

It is easy to verify that \eqref{eq:g-BF1} with $\Sigma_{\gamma \gamma}=\overline{X}^t \overline{X}/n$ yields the $g$-Zellner BF. 
Notice that this expression generalizes Jeffreys' proposal; however, to the best of our knowledge, it is not a generalization of Zellner's proposals.

\section{Discussing priors for the general case}\label{app.priors}
Without loss of generality, the prior for each model $\n\gamma$ can be expressed as
$$
\pi_\gamma(\n\alpha,\n\beta_\gamma,\n\nu)=
\pi_\gamma(\n\nu)\,\pi_\gamma(\n\alpha\mid\n\nu)\pi_\gamma(\n\beta_\gamma\mid\n\alpha,\n\nu).
$$

\paragraph*{About $\pi_\gamma(\n\nu)$} The prior distribution over the parameters of the imputation model, denoted as $\pi(\n\nu)$, is a common component across all models. This is because, as previously mentioned, all the covariates are incorporated into this model. Without additional information, the recommendation is to use a prior endorsed by the literature on objective estimation priors (the reference prior, if available), provided that the corresponding posterior distribution is proper.

\paragraph*{About $\pi_\gamma(\n\alpha\mid\n\nu)$} The parameters $\n\alpha$ are common to all the models considered but are of a different nature from $\n\nu$. The prior $\pi_\gamma(\n\alpha\mid\n\nu)$ appears in the marginal $m_\gamma({\n y}_{(0)}, {\n x}_{(0)})$ as a multiplicative function in $\mathfrak{m}_\gamma({\n y}_{(0)}\mid {\n x}_{(0)}, {\n x}_{(1)}, \n\nu)$ (see Equation~\ref{cal.mar}). This implies that the undetermined proportionality constant in improper prior transfers to $m_\gamma({\n y}_{(0)}, \widetilde{\n x}_{(0)})$, which is automatically undetermined. Fortunately, what matters in computing posterior probabilities is not the marginal itself but the ratio of the two (the BF). This opens a possible justification for the use of improper priors. This is because if the same priors are used for all models (recall that $\pi_\gamma(\n\alpha\mid\n\nu)$ appears in all models), then the undetermined constant would cancel out, as it does in the posterior distribution in the estimation settings. The argument is debatable. However, this can easily be accompanied by more convincing limiting arguments in which the (common) improper prior is expressed as a limit of a proper prior density with a well-defined proportionality constant that cancels out. Furthermore, because it is an objective prior, the dependence on $\n\nu$ is irrelevant, leading to $\pi_\gamma(\n\alpha\mid\n\nu)=\pi_\gamma^N(\n\alpha)$

The question now is when it is justified to use the same prior $\pi_\gamma^N(\n\alpha)$ for all models, and, of course, which to use. The short answer is that it is reasonably justified if $\n\alpha$ has a similar interpretation in all models; in this case, we should use an objective estimation prior. The standard practice assumes that a similar meaning holds when the common parameters are orthogonal to the new ones; this is why the design matrix is expressed in terms of the mean in the regression (which is not possible here).

Arguments regarding common parameters and orthogonality have accompanied the development of BF since their conception \citep{Jeff:61}. However, these are far from formal, and several authors have opted for other approaches to the problem. This is the case in \cite{Berg:Peri:01}, who expressed the possibility of using predictive matching arguments to develop more formal arguments for handling prior assignments for common parameters. Their work is made more explicit in \cite{Berg:Peri:Vars:98} and reviewed in a more general setting in \cite{Baetal11}. 

\paragraph*{About $\pi_\gamma(\n\beta_\gamma\mid\n\alpha,\n\nu)$} This is perhaps the most delicate ingredient in the prior assignment. Similarly to the discussion of $\pi_\gamma(\n\alpha\mid\n\nu)$ above, the prior for $\n\beta_\gamma$ enters multiplicatively into the equation for $\mathfrak{m}_\gamma({\n y}_{(0)}\mid {\n x}_{(0)},{\n x}_{(1)}, \n\nu)$ and, if an improper prior is used, its undetermined constant is transferred directly to the marginal. However, the parameter $\n\beta_\gamma$ is specific to $\n\gamma$, and thus, the possibility of canceling out the constants disappears. Hence, the prior $\pi_\gamma(\n\beta_\gamma\mid \n\alpha,\n\nu)$ must be proper, a requirement that cannot be circumvented by a ``proper'' vague prior (which would hide the problem, not solve it).

Seminal works in this area include a series of papers \cite{Zell:Siow:80, ZellSiow84} and \cite{Zell:86} that introduced the popular $g$-priors in the context of normal regression models. The $g$-prior approach uses a zero-mean multivariate normal distribution for $\n\beta_\gamma$ with a covariance matrix obtained from the expected Fisher information matrix. Many popular proposals in the literature are generalizations of this basic idea. The domain of normal linear models includes benchmark priors \citep{FLS01}, hyper-$g$-priors \citep{liang08}, robust priors \citep{Baetal11}, generalised linear models \citep{LiClyde18, held2015approximate, Saba:Held:11}, and survival models \citep{a-2022biometrics,a-2020bayesiananalysis}.

\section{Further experiments}
\subsection*{Experiment S1: Model selection}
We simulate three variables $y,x_1,x_2$ with the following scheme:
$$
y_i=1+\beta_1^{True} x_{i1}+\beta_2^{True} x_{i2}+N(0,1),\,\, 
\left(\begin{array}{c}
	x_{i1} \\ x_{i2} 
\end{array}\right)\diid N_2
\left( \left( \begin{array}{c} 1 \\ 2 \end{array}\right),
\left(\begin{array}{cc} 1 & \rho^{True} \\ \rho^{True} & 1 \end{array}\right)\right).
$$
We are interested in the model choice problem with competing models
$$
f_0(\n y\mid \n x_1,\n x_2,\alpha,\sigma)=N_n(\n y\mid \alpha\n 1, \sigma^2I),\,\,
f_1(\n y\mid \n x_1,\n x_2,\alpha,\sigma)=N_n(\n y\mid \alpha\n 1+X_1\beta_1, \sigma^2I).
$$
Broadly, we are interested in testing whether $y$ is affected by $x_1$ and $x_2$ plays the role of an extra variable in our dataset. Here, $f_0$ represents the null model (no association) and $f_1$ is the alternative model (association). Although covariate $x_2$ is not directly involved in the testing problem, it is involved in the imputation process. 

We simulated 6000 datasets, each with a sample size of $n=100$, following the above probabilistic structure but with several parameter configurations. In particular, we consider four different levels of correlation, $\rho^{True}\in\{0, 0.4, 0.7, 0.9\}$ and three different scenarios:
\begin{equation*}
	S1:\beta_1^{True}=0.3,\beta_2^{True}=0; \quad S2:\beta_1^{True}=\beta_2^{True}=0; \quad S3:\beta_1^{True}=0,\beta_2^{True}=0.3.
\end{equation*}
In this case, the highest signal-to-noise ratio value is approximately 0.3, significantly lower than the design's value of 1.26 in Experiment~1.

In S1, $y$ relates to $x_1$ (the alternative model is true). By contrast, in S2, there is no relationship with $x_1$  (the null model is true), either directly or indirectly through $x_2$. In S3, $y$ is correlated only with $x_2$; thus, its relationship with $x_1$ is indirect (as the correlation between $x_1$ and $x_2$ increases, we move from the situation where the true model is $f_0$ to the situation where the true model is $f_1$). Finally, to bring the issue of missingness into the picture, a certain proportion $\pi\in\{0.05, 0.15, 0.40, 0.60, 0.75\}$ of the values of variable $x_1$ are missing. The values are amputated using the MAR and MCAR mechanisms, resulting in an incomplete dataset $\widetilde{\n y}, \widetilde{\n x}_{(0)}$.

\begin{figure}[t!]
	\begin{center}
		\begin{tabular}{ccc}
			\includegraphics[width=5cm, height=12cm]{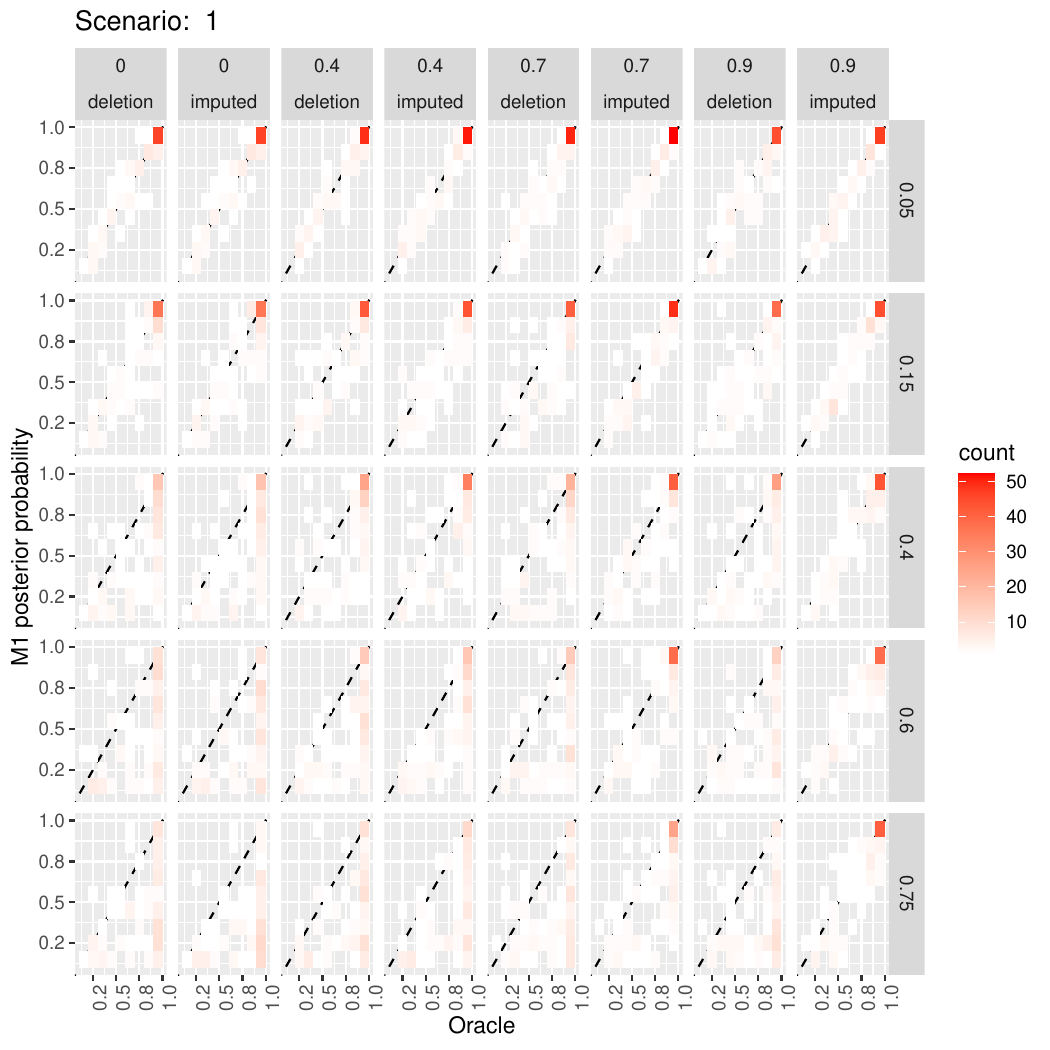} &
			\includegraphics[width=5cm, height=12cm]{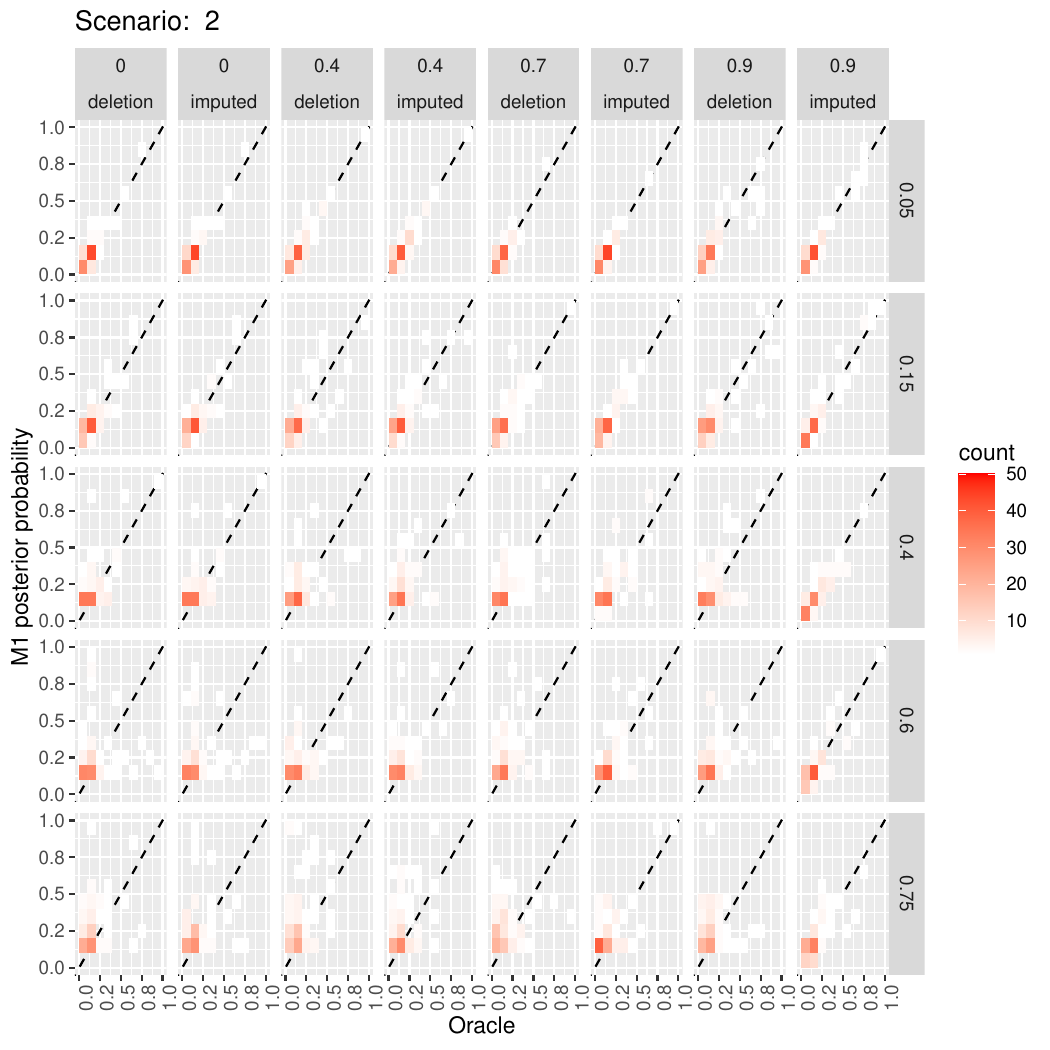} &
			\includegraphics[width=5cm, height=12cm]{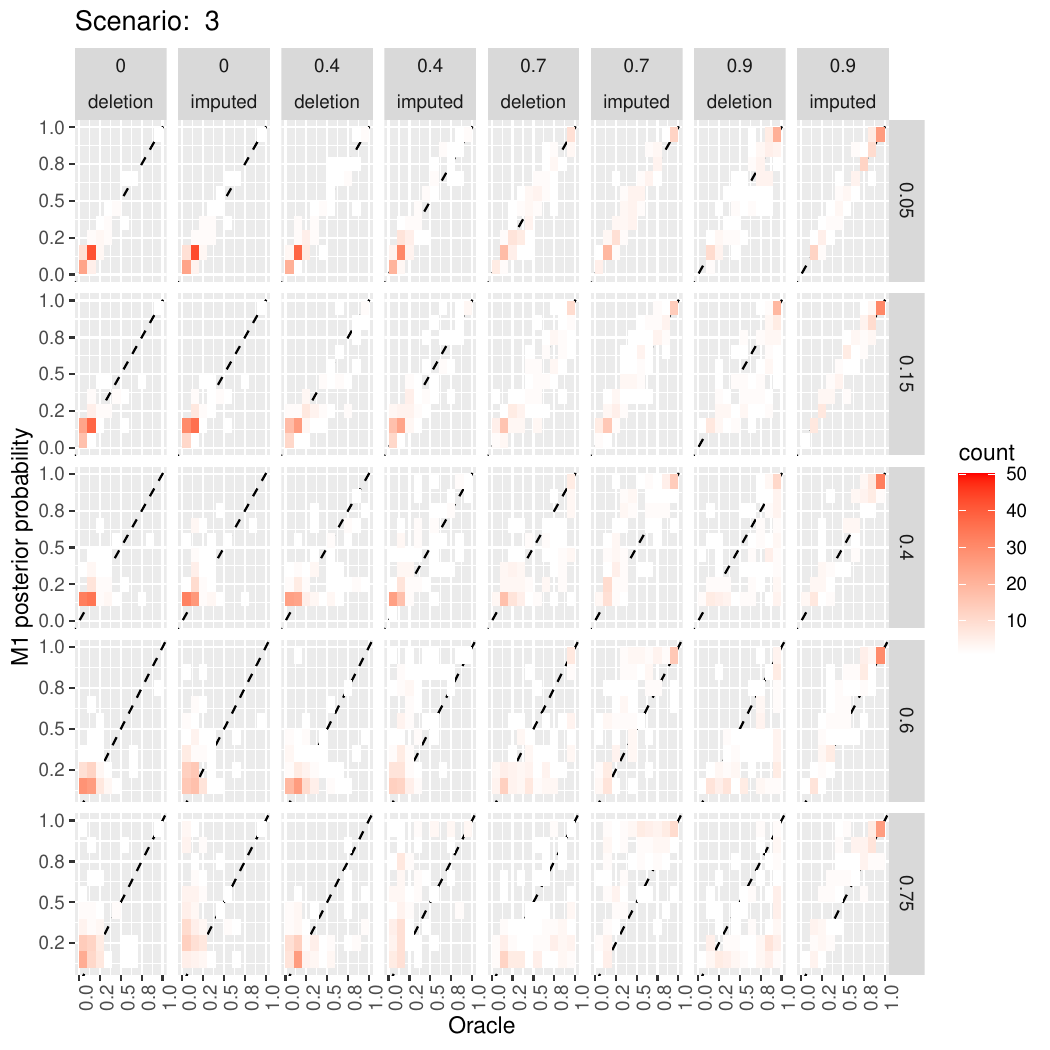}
		\end{tabular}
	\end{center}
	\caption{Experiment S1. Bivariate (oracle vs. deletion and oracle vs. imputation) absolute frequencies of posterior probabilities in favor of the alternative model for S1 (left), S2 (middle), and S3 (right). The results are under MAR and each row of graphs represents a proportion of missing data, with the lowest percentage at the top. The results for the different values of $\rho^{True}$ are in columns, alternating results for deletion with those for imputation, with the last two columns corresponding to the highest correlation.}\label{Ex1}
\end{figure}

Next, $N=100$ datasets were generated for all combinations, leading to the above $100 \times 3 \times 4 \times 5=6000$ datasets being considered for each missing data mechanism. For each dataset, we compute the posterior probabilities in favor of the alternative model based on the oracle $g$-BF, listwise deletion $g$-BF, and imputed $g$-BF. The MAR mechanism results are summarised in the form of image panel plots (see Figure~\ref{Ex1}), one for each scenario, representing the frequency of the bivariate posterior probabilities (imputed vs. oracle; listwise deletion vs. oracle). Similar results were obtained with MCAR (data not shown). The resulting plots are easy to interpret; the darker the colors on the diagonal, the better the evidence retained after missingness.

Regarding the effect of losing observations in general and comparing the three sce\-na\-rios, we observed that the consequences of missingness were clearly worse when the alternative model was true than when the null model was true. In other words, missingness affects sensitivity more than specificity as agreements with the oracle are less frequent in S1 than in S2 and in the right columns of S3 than on the left. Furthermore, S1 shows that sensitivity is reduced, whereas specificity becomes more variable.

Comparing the behavior of the two approaches in S1, we appreciate the superiority of imputation in preserving evidence in favor of the alternative model. As expected, this improved performance became more pronounced as the proportion of missing values and/or $\rho^{True}$ increased. In S2, imputation and deletion behaved similarly for small-to-moderate correlations, with imputation performing slightly better at $\rho^{True}=0.7$. For a very high correlation between covariates (right columns), the imputed $g$-BF is clearly preferable, regardless of the proportion of NA observations. The similarities in the procedures were maintained in S3 for moderate correlations ($\rho^{True}\le.4$) and a proportion of missingness $\le 40\%$. 

However, interesting features emerge for both when the proportion of missingness is high ($\ge 60\%$), where deletion preserves the oracle responses slightly better, which provides little evidence for the null model. By contrast, imputation recovers oracle responses much better when the alternative model receives higher probabilities. The plots corresponding to $\rho^{True}=0.7$ and a proportion of missingness of 0.75 (or 0.6) are particularly interesting, as imputation and deletion behave differently. We observed that deletion responds to missingness by lowering oracle responses, similar to what we observed in S1. Simultaneously, imputation tends to subtly increase the strength of the signal, owing to the influence of $x_2$, which is used for imputation. This effect is diluted when the correlation is very high because the original and imputed variables are similar.

The above findings confirm the conclusions of Experiment 1 and seem to align with commonsense intuition: imputation is preferable to elimination when $\rho^{True}$ and/or the percentage of missing data is higher. However, this effect and the improvement in specificity are more noticeable in the context of Experiment~1 (variable selection), which allows us to conclude that imputation gains strength when there are several covariates from which to learn. In contrast, the e\-li\-mi\-nation method is more at risk because it is more likely to lose a larger sample size.


\subsection*{Experiment S2. The Boston dataset}

In this section, we describe an experiment similar to the previous one using the Boston dataset first analyzed in \cite{Harrison:1978}, which is available in the R package {\tt MASS} \citep{MASS-book-2002}. In contrast to the Ozone dataset, the normality assumption was less reasonable for some of the variables in the Boston dataset. 
\begin{figure}[h]
	\begin{center}
		\includegraphics[width=10cm,height=8cm]{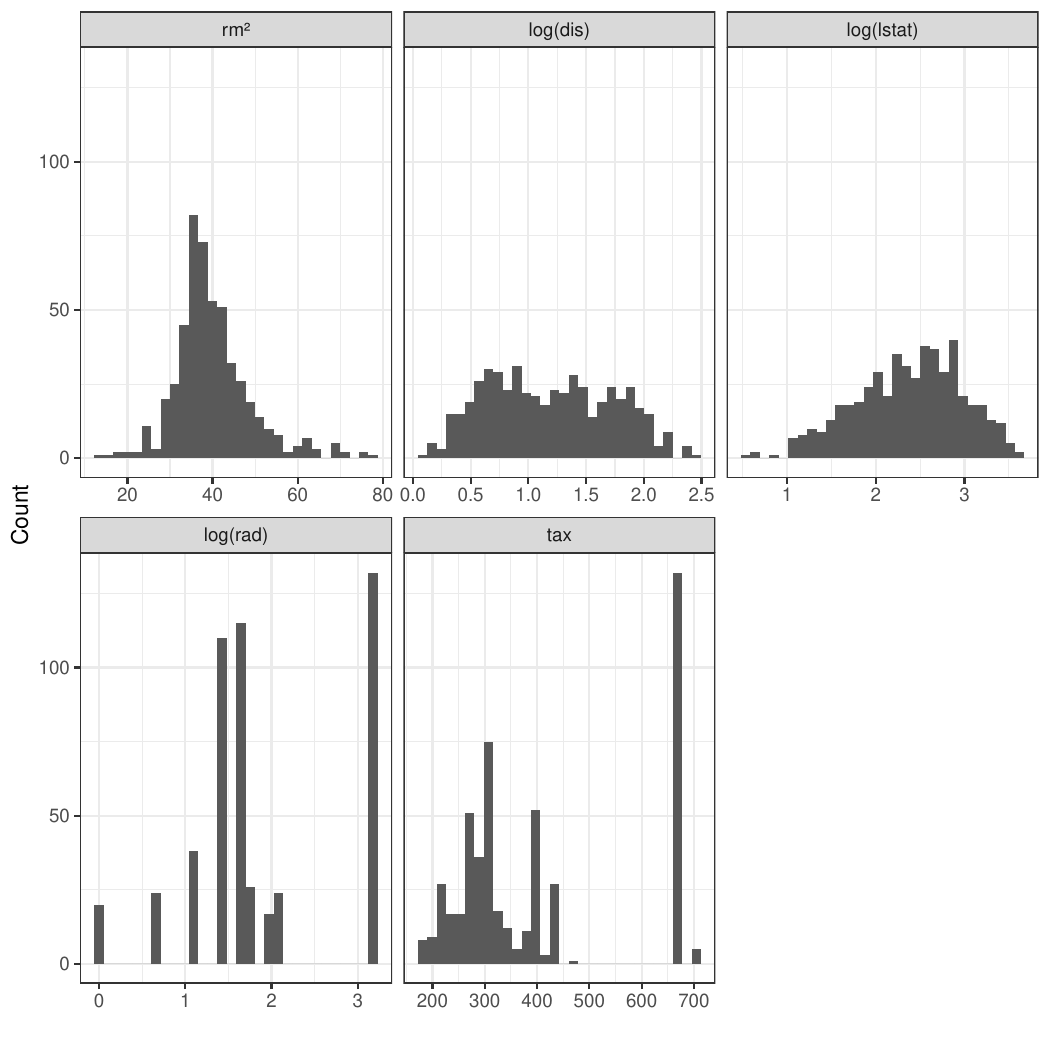} 
	\end{center}
	\caption{Boston dataset. Histograms of variables in which missing values are introduced.} \label{Boston.descri}
\end{figure}
The Figure~\ref{Boston.descri} of the supplementary material contains histograms of the variables where missing data were introduced.

The response variable was the median of owner-occupied dwellings in the census on a logarithmic scale. There were 13 covariates, some of which were transformed (logarithmic or quadratic) as in the original study. The dataset contains $n=506$ observations and we use an MAR mechanism to remove observations from the variables {\tt rm2}$_\star$, {\tt log.dis}$_\star$, {\tt log.rad}$_\star$, {\tt tax}$_\star$ and {\tt log.lstat}$_\star$ (the symbol $\star$ explicitly indicates that these variables have NA va\-lues). The probability of missing each variable depends on all remaining fully observed covariates. In Figure~\ref{Boston.descri}
we see that {\tt rm2}$_\star$, {\tt log.dis}$_\star$ and {\tt log.lstat}$_\star$ have an empirical distribution that can be reasonably well modeled by a normal, while {\tt log.rad}$_\star$ and {\tt tax}$_\star$ show strong bimodality, making the normality hypothesis inadequate. This acts against the imputed $g$-BF. Part of our interest in this dataset was to check how this procedure behaves under strong departures from the normality of variables with missing data. The percentages of missing values per variable were 10, 20, and 30\%, resulting in mean overall missing values per\-cen\-ta\-ge of 35, 57, and 72\%, respectively. As in the previous experiments, 100 replicates were performed for each missing percentage. 

\begin{figure}
	\begin{center}
		\includegraphics[width=8cm, height=7.5cm]{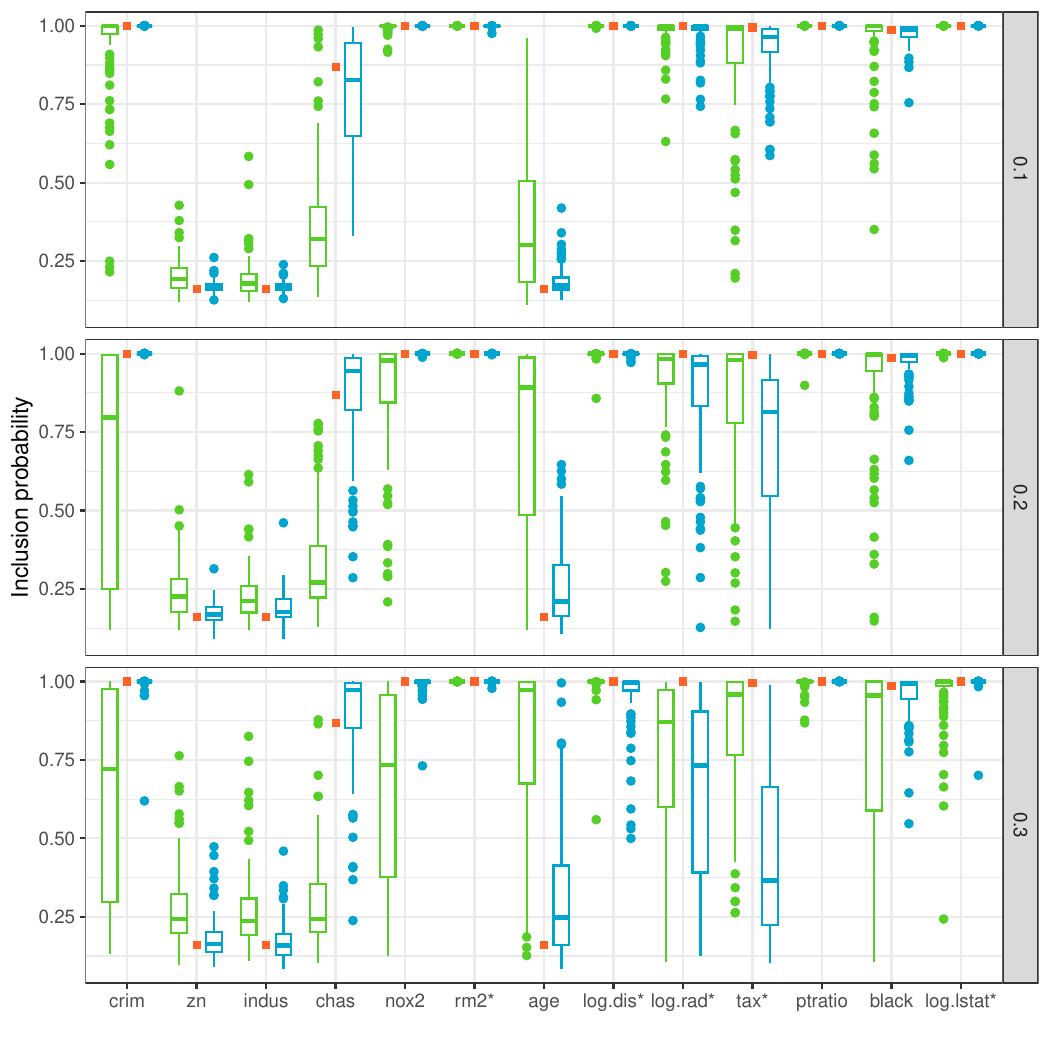}
	\end{center}
	\caption{Boxplots of the inclusion probabilities for each variable using imputed $g$-BF (blue) and listwise deletion $g$-BF (green), when considering 10 (top), 20 (middle) or 30\% (bottom) of missing values per variable, for Boston dataset. The corresponding oracle $g$-BF inclusion probabilities are depicted in red. The symbol $\star$ in variable names explicitly indicates the variables with missing data.} 
	\label{all.bp.Boston}
\end{figure}
The results of the posterior inclusion probabilities are summarized in Figure~\ref{all.bp.Boston}. 
As in previous experiments, the deletion results were substantially more dispersed than those of imputation, which was more self-controlled because of the difference in the final sampling information used by the two methods. Furthermore, it is important to conclude that imputation has clear benefits for most variables not affected by missingness: {\tt crim}, {\tt zn}, {\tt indus}, {\tt nox2}, {\tt age}, {\tt ptratio}, and {\tt black}. The other variable in this category, {\tt chas}, requires additional discussion, which is addressed next. 

Among the amputated variables, deletion moderately outperforms imputation for those that violate Gaussianity---{\tt log.rad}$_\star$ and {\tt tax}$_\star$---, whereas the remaining imputation produces results closer to the oracle---{\tt log.lstat}$_\star$---or behaves similarly---{\tt rm2}$_\star$ and {\tt log.dis}$_\star$---. In the case of {\tt chas}, listwise deletion loses all signals, even for 10\% of the missing data. Simultaneously, imputation somewhat increases the evidence in favor of this variable, especially for the highest percentage of missing data, which may result from losing e\-vi\-den\-ce on other imputed variables, with a byproduct of partial correlation explaining the response.

\subsection*{Experiment S3. Comparing distributions of the errors} 
To illustrate the methods discussed in Section~\ref{sec.errors}, particularly the case treated in Example~\ref{example.errors}, we designed a simulated experiment inspired by the {\tt advertising} dataset of \cite{James2013}, Chapter 2, which focuses on the issue of the possible heteroscedasticity of errors. The data consists of product sales (dependent variable) and product advertising budgets for $p=3$ different media (TV, radio, and newspaper) from 100 different markets. 
\begin{figure}
	\begin{center}
		\includegraphics[scale=0.28]{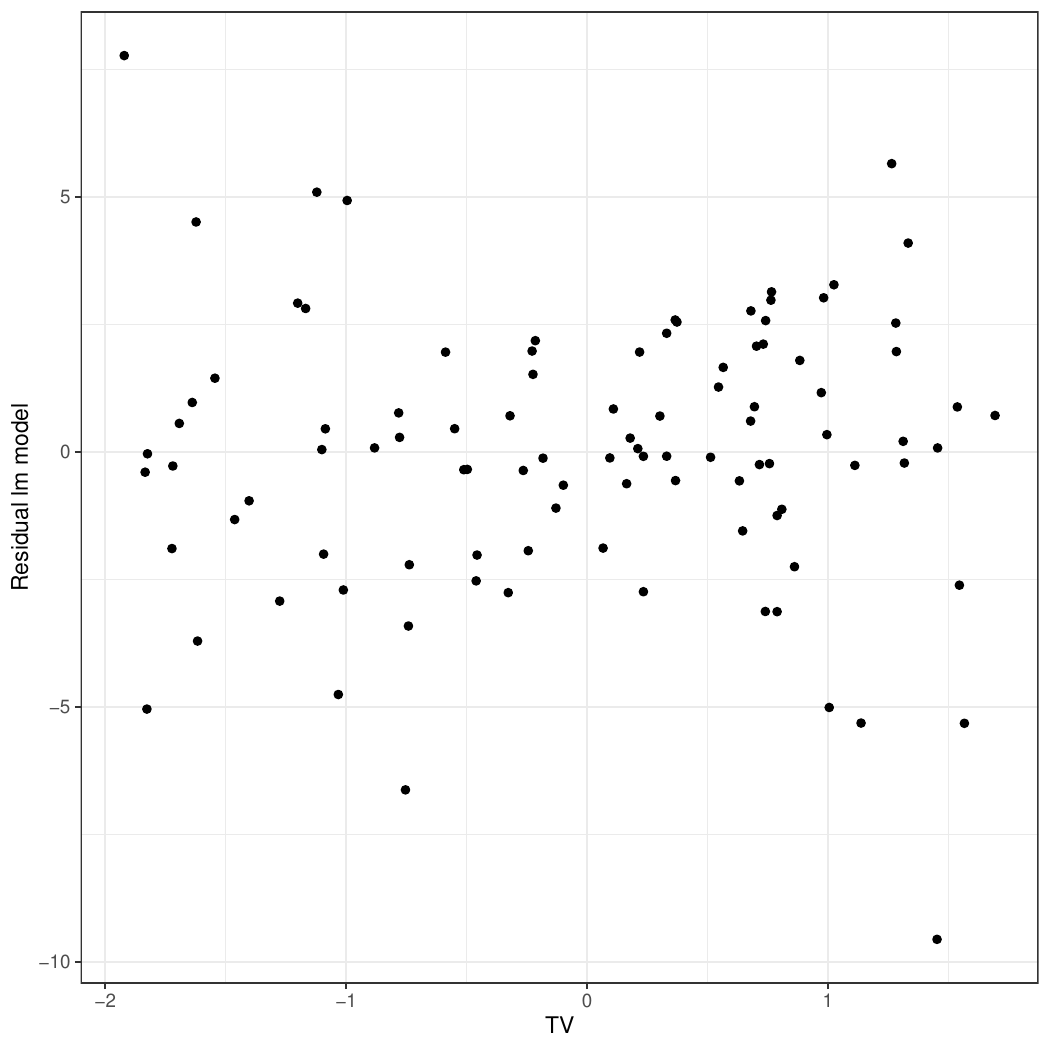} 
	\end{center}
	\caption{Residuals plot from the linear model {\tt sales $\sim$ TV + radio} against {\tt TV}.} \label{Residual.advert}
\end{figure}
We consider testing homoscedasticity against a given form of heteroscedasticity. More precisely, we test the error distribution in Equation~\eqref{distrib.errors} with $\Psi=\mbox{diag}(\sqrt{|TV_i|})$; that is, the error variance depends (via the squared root) on TV budget. The code used to obtain this dataset, generated under this hypothesis, is in the above-mentioned GitHub location. The residuals of the considered model with respect to the values of TV in the simulated data are shown in Figure~\ref{Residual.advert}.

\begin{figure}
	\begin{center}
		\includegraphics[scale=0.4]{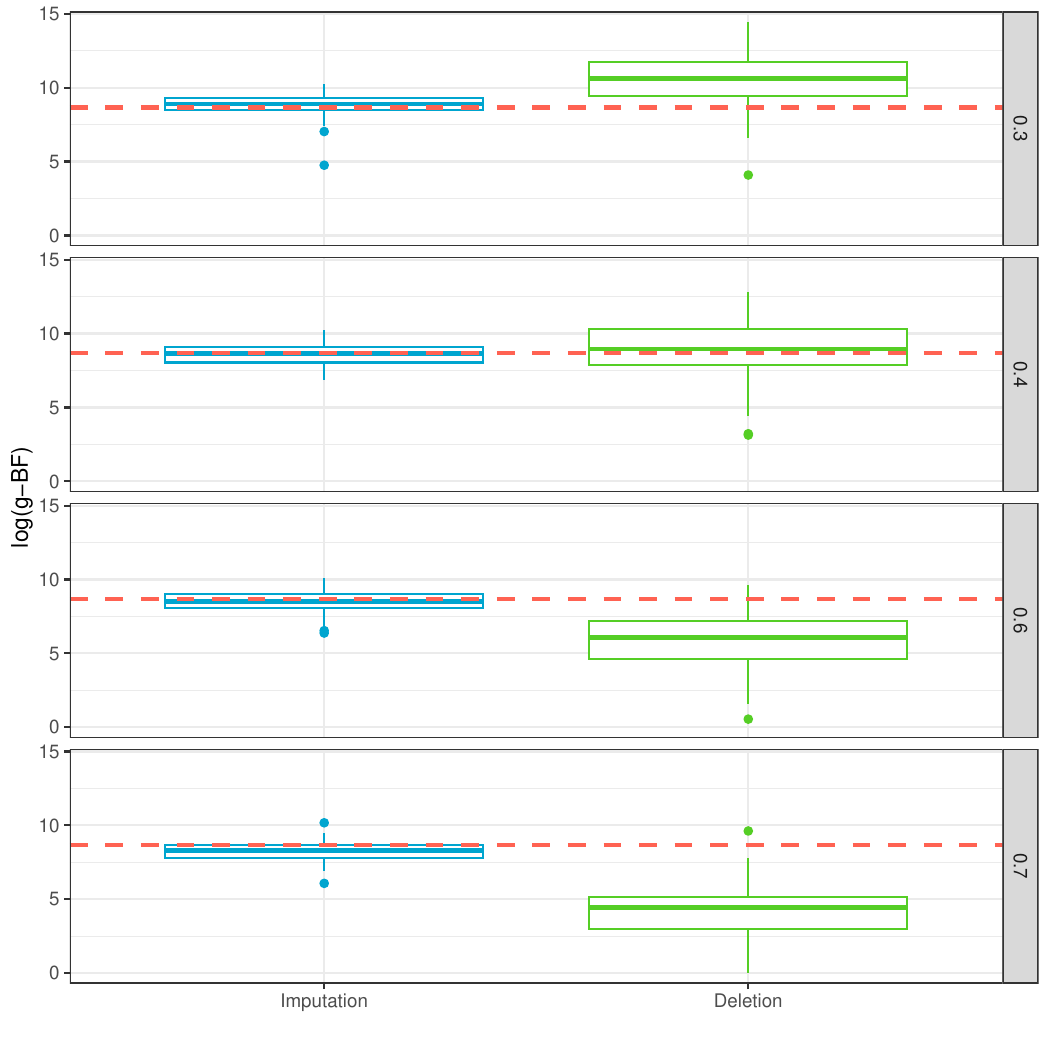} 
	\end{center}
	\caption{Experiment S3. Boxplots of log(BF) comparing imputation (blue) with listwise deletion (green) method and the oracle (red) for 30 (top), 40, 60, and 70\% (bottom) of missing data.} \label{Boxplot-advert-sim}
\end{figure}

As in the preceding experiments, we induced some NA values in {\tt radio}, with 30, 40, 60\%, and 70\% missing values, assuming MAR (through the {\tt ampute} function of the {\tt mice} package, with missingness depending on the values of {\tt newspaper}) and MCAR. In Figure \ref{Boxplot-advert-sim}, we present the results for the MAR mechanism (the results for MCAR are similar) by comparing our proposal of an imputation log(BF) (cf. \eqref{eq:g-BF2}) with the oracle (same equation but with the full dataset $X_1$, so the expectation does not have any effect) and listwise deletion (same as oracle but with $X_1$ only containing the full observed rows). Although the two methods correctly choose the model (even for the highest percentage of missing data), we observe greater variability in the deletion-based method, which is accompanied by a tendency to dilute the evidence initially reported by oracle BF. By contrast, the imputed BF remains closer to the oracle, with less variability at all levels of missingness.

\bibliography{merged.bib}
\bibliographystyle{chicago}

\end{document}